\documentstyle[preprint,eqsecnum,aps,prabib]{revtex}

\begin{document}

\draft

{\tightenlines

\title{Comparison of triton bound state properties using different
separable representations of realistic potentials}

\author{W.~Schadow}
\address{Institute of Nuclear and
Particle Physics, and Department of Physics, Ohio University,
Athens, OH 45701}

\author{W.~Sandhas}
\address{Physikalisches Institut der Universit\"at Bonn, Endenicher
Allee 11-13, D-53115 Bonn, Germany}

\author{J.~Haidenbauer}
 \address{Institut f\"ur Kernphysik, Forschungszentrum J\"ulich GmbH,
D-52425 J\"ulich, Germany}

\author{ A.~Nogga}
\address{Institut f\"ur Theoretische Physik II, Ruhr-Universit\"at Bochum,
D-44780 Bochum, Germany}

\date{October 22, 1998}

\maketitle

\begin{abstract} The quality of two different separable expansion
methods ({\sl W} matrix and Ernst-Shakin-Thaler) is investigated. We
compare the triton binding energies and components of the triton wave
functions obtained in this way with the results of a direct
two-dimensional treatment.  The Paris, Bonn {\sl A} and Bonn {\sl B}
potentials are employed as underlying two-body interactions, their total
angular momenta being incorporated up to $j \leq 2$.  It is found that
the most accurate results based on the Ernst-Shakin-Thaler method agree
within 1.5\% or better with the two-dimensional calculations, whereas
the results for the {\sl W}-matrix representation are less accurate.
\end{abstract}

\pacs{PACS number(s): 21.60.-n, 27.10.+h }
}

\narrowtext

\section{Introduction}

Separable approximations or expansions of the underlying two-body
interaction play an essential role in calculations of few-nucleon
systems.  In the three-body problem such an input reduces the
two-dimensional Faddeev equations, or the corresponding
Alt-Grassberger-Sandhas (AGS) equations, to one-dimensional effective
two-body equations~\cite{Alt67}. Analogously, the four-body AGS
equations go over into effective three-body, and, after repeated
application of separable expansions, into effective two-body
equations~\cite{Grassberger67a}. A considerable reduction of the
complexity of the original problem is achieved in this way.  What
remains to be done, however, is a careful test of accuracy of the
respective separable representations.

In early calculations Yamaguchi-type separable potentials were
employed. Though not very realistic, they simulated characteristic
aspects of the nuclear interaction, leading thus to qualitatively
correct cross sections for three-nucleon scattering.  Stimulated by this
experience, rather sophisticated separable approximation or expansion
techniques were developed and have been applied successfully to
realistic interactions.

The Ernst-Shakin-Thaler (EST) expansion \cite{Erns73} is considered as a
particularly powerful separable approach.  Its efficiency has been
thoroughly investigated for the Paris potential \cite{Haid84}. Indeed,
based on high-rank separable expansions of this type
\cite{Haid86a,Koike87a}, rather accurate predictions for the
neutron-deuteron ($n$-$d$) cross section and some polarization
observables were achieved \cite{Koike87a,Koike87b}. Later on, when
direct solutions of the two-dimensional three-body equations became
available \cite{Wital88a}, it was established that the results obtained
in this way and with the EST approach are in excellent agreement
\cite{Corne90a,Nemoto98a}.

The {\sl W}-matrix method \cite{Bart86} provides quite directly a
rank-one separable representation of the two-body {\sl T} matrix, which
preserves the analytical properties of the original {\sl T} matrix. For
the Malfliet-Tjon (MT I+III) potential three-body bound-state and
scattering results \cite{Bart87,Janu93}, including break-up
\cite{Frank88b}, were obtained on this basis in complete agreement with
alternative treatments.  Calculations for the Paris potential were also
found to be rather reliable for most, but not all observables
\cite{Janu93}.

For further applications of separable approximations or expansions,
e.g., in processes as the photodisintegration of three-nucleon systems
\cite{Schad97a,Sandhas98a}, the corresponding radiative capture
\cite{Schadow98c}, or the pion absorption on such systems
\cite{Cant97a}, additional tests of accuracy of the three-body wave
functions involved appear desirable. In case of a 5 channel
calculation and the EST method such a test has been made already in
Ref. \cite{Park91} for the Paris potential, and, with a modified
expansion method, in Ref. \cite{Koike97a} for the Argonne 14 potential
restricted to $j \leq 2$. 

In what follows we compare triton bound-state calculations performed
by means of the EST expansion and the {\sl W}-matrix approximation
with direct solutions of the two-dimensional homogeneous three-body
equations. In view of the sensitivity of the bound-state problem to
differences in the two-body input, this comparison should provide a
particularly relevant test of accuracy of these methods.

The paper is organized as follows: Sections \ref{secwmatrix} and
\ref{secestmethod} briefly describe the {\sl W}-matrix and EST
methods, respectively. In Sec. \ref{secbound} we give the relevant
equations for the bound-state calculations using separable potentials
(details of the direct treatment can be found in
Refs. \cite{Gloeck83,Nogga97a}).  Our results and conclusions are
presented in the last section.

\section{{\sl W}-matrix representation}
\label{secwmatrix}

The {\sl W}-matrix method \cite{Bart86,Janu93} leads to a quite
appropriate splitting of the two-body {\sl T} matrix into one dominant
separable part and a small non-separable remainder

\begin{equation}
T_{l l'}(p, p'; E + i0) = {\rm {^S}}T_{l l'}(p, p'; E + i0)
+ {\rm {^R}}T_{l l'}(p, p'; E + i0) .
\label{tmatsplit}
\end{equation}

Similarly to the {\sl T} matrix, the {\sl W} matrix is defined by an
equation of Lippmann-Schwinger type, however, with a modified
non-singular kernel.  After partial wave decomposition this equation is
given by (we use units $(\hbar c)^2 = 2 \mu = 1$ and thus $ E = p^2$)

\begin{equation}
 W_{l \hat l}^\eta (p, p'; E) \, = \, U^\eta_{l \hat l}(p,p') + \sum_{l'}
 \int\limits^\infty_0 dq \, q^2
 \, {{U^\eta_{l l'} (p,q) - U^\eta_{l l'} (p, k)} \over {E - q^2}} \, q^{2 l'}
 \,  W^\eta_{l' \hat l} (q, p'; E),
\label{wmat}
\end{equation}

\noindent where $k$ is subject to the constraint $k = \sqrt{E}$ for $
E \geq 0 $ and is kept arbitrary for $E < 0$.  Here $l$ and $\hat l$
are the orbital angular momenta, and $\eta = (s,j;t)$ stands for the
spin, the angular momentum j [ with the coupling sequence $(l,s)j$],
and the isospin $t$ of the two-body subsystem.  The input entering Eq.
(\ref{wmat}) is related to the two-body potential matrix $V^\eta_{l
\hat l}(p,p')$ according to $ U^\eta_{l \hat l} (p,p') = p^{-l} \,
V^\eta_{l \hat l} (p,p') \, p'^{\hat l}$. The separable part of the
{\sl W}-matrix representation (\ref{tmatsplit}) of the two-body {\sl
T} matrix is given by

\begin{equation}
  {\rm {^S}}T^\eta_{ll'} (p,p'; E + i0) =
 \sum_{\hat l \hat l'} p^l\, W^\eta_{l \hat l} (p, k; E) \,
\Delta^\eta_{\hat l \hat l'} (E + i0) \, W^\eta_{l' \hat l'} (p', k; E) \,
 p'^{l'}
\label{tmatsepw}
\end{equation}

\noindent
and the nonseparable remainder reads

\begin{equation}
  {\rm {^R}}T^\eta_{ll'} (p,p'; E + i0) = p^l  \left [
W^\eta_{l  l'} (p, p'; E) -
 \sum_{\hat l \hat l'}  W^\eta_{l \hat l} (p, k; E) \,
(W^\eta_{\hat l \hat l'} (k, k; E))^{-1} \,
 W^\eta_{ \hat l' l'} (p', k; E) \right ]
 p'^{l'} .
\label{tmatsepr}
\end{equation}

\noindent
The propagator $\Delta^{\eta}_{\hat l \hat l'}$ is given by

\begin{equation}
 \Delta^{\eta}_{\hat l \hat l'} = \sum_{\hat l''}
(F^{\eta}_{\hat l \hat l''}(E+i0))^{-1} \,
(W^{\eta}_{\hat l \hat l''}(k, k; E))^{-1} ,
\end{equation}

\noindent
where $F^{\eta}_{\hat l \hat l''}(E+i0)$ is a generalization of the
Jost function

\begin{equation}
(F^{\eta}_{\hat l \hat l'}(E+i0))^{-1} = \delta_{\hat l \hat l'}
- \int\limits_{0}^{\infty} dq \, q^{2} \, q^{2 \hat l} \,
\frac{W_{\hat l \hat l'}(q, k; E)}{E + i0 - q^{2}} \,.
\end{equation}

On-the-energy-shell and half-off-shell the separable part
(\ref{tmatsepw}) of (\ref{tmatsplit}) is identical with the exact {\sl
T} matrix. In fact, when inserting the momentum $k = \sqrt{E}$ for one
or both of the momenta $p$ or $p'$ , we easily see that the remainder
(\ref{tmatsepr}) vanishes. Therefore, the separable part of the {\sl
T} matrix has the same pole and cut structure as the full {\sl T}
matrix. This suggests to approximate the $T$ matrices, entering the
kernel of Faddeev-type equations, by the separable expression
(\ref{tmatsepw}).  To optimize this approximation, i.e., to minimize
the effect of the neglected remainder, two criteria for the choice of
the functional form of the free parameter $k$ were developed
\cite{Janu93}.  In the present work we apply the criterion based on
the Schmidt norm of the kernel of the three-body equations.

In the past this method has been used in elastic $n$-$d$ scattering
calculations \cite{Bart87,Janu93}, in the breakup case \cite{Frank88b},
and more recently in the photodisintegration of the triton
\cite{Schad97a}.

\section{EST method}
\label{secestmethod}

The Ernst-Shakin-Thaler (EST) method \cite{Erns73} allows one to
generate separable representations of arbitrary rank $N$ that agree
exactly (on- and half-off-shell) with the original $T$ matrix at $N$
specific, appropriately chosen energies.

For a brief outline of this method let us begin with the (partial-wave
projected) Lippmann-Schwinger equation for the wave function,

\begin{equation}
|\psi_E \rangle = |k_E\rangle + G_0 (E)
V |\psi_E\rangle,
\label{est1}
\end{equation}

\noindent where $|k_E\rangle$ is the incoming wave, and $G_0(E)$ the
two-body Green's function. The dependence on the orbital angular momenta
$l$, $l'$ and on the conserved quantum numbers (the spin, the angular
momentum j [ with the coupling sequence $(l,s)j$], and the isospin $t$
of the two-body subsystem) is suppressed for convenience.  For proper
scattering solutions (on-shell), $k_E$ and $E$ are related by $ E = k^2_E
$.

According to the EST method, a rank-$N$ separable representation of a
potential $V$ is given by the form

\begin{equation}
 V^{\rm EST} = \sum_{\mu,\nu=1}^N V |\psi_{E_\mu} \rangle \,
\Lambda_{\mu \nu} \, \langle \psi_{E_\nu} | V  ,
\label{est2}
\end{equation}

\noindent where $E_\mu$, ($\mu=1,...,N$), is a freely choosable, but
fixed set of energies.  The coupling strengths $\Lambda_{\mu \nu}$ are
determined by the condition

\begin{equation}
\sum_{\nu=1}^N \Lambda_{\mu \nu} \langle \psi_{E_\nu} | V
|\psi_{E_\rho} \rangle = \delta_{\mu \rho}  .
\label{est3}
\end{equation}

\noindent Note that the "form factors" in the separable potential
(\ref{est2}) consist of the objects $V|\psi_{E_\mu} \rangle$, where
$|\psi_{E_\mu}\rangle $ are solutions of Eq. (\ref{est1}) at the
energies $E = E_\mu$. Thus, Eq. (\ref{est3}) together with
Eq. (\ref{est2}) implies that the following relation holds at the $N$
energies $E_\mu$

\begin{equation}
 V^{\rm EST} |\psi_{E_\mu} \rangle = V |\psi_{E_\mu} \rangle =
 T (E_\mu) |k_{E_\mu} \rangle =   T^{\rm EST} (E_\mu) |k_{E_\mu} \rangle .
\label{est4}
\end{equation}

\noindent Here $T$ and $T^{\rm EST}$ are the two-body $T$ matrices for
the potential $V$ and its separable representation $ V^{\rm EST}$,
respectively. Evidently Eq. (\ref{est4}) means that the on-shell, as well
as the half-off-shell $T$ matrices for both interactions, $V$ and
$V^{\rm EST}$, are exactly the same at the energies $E_\mu$.

With the form factors $| g_{\nu}^{\eta} l \rangle =
V^\eta|\psi^\eta_{E_\nu} \rangle_l$ ($\eta = (s,j;t)$) and the
propagators ${ \Delta}_{ \mu \nu}^{\eta}$ of this representation, the
two-body {\sl T} matrix reads

\begin{equation}
\label{eqtmatest}
T^{\eta}_{l l'}(E +i0)  = \sum_{\mu \nu }| { g}_{\mu}^{\eta}  l\rangle \,
{ \Delta}_{\mu \nu }^{\eta} (E+i0) \,\langle { g}_{\nu}^{\eta} l'|
\end{equation}

\noindent
where

\begin{equation}
 {\bf \Delta}^{\eta} (E + i0) = \left (({{\bf \Lambda}^{\eta}})^{-1} -
{\bf {\cal G}_{0}}(E + i0) \right )^{-1}
\end{equation}

\noindent
and

\begin{equation}
({\bf {\cal G}}_{0}(E + i0))_{\mu \nu} = \sum_l
\langle g_{\mu} l |G_0(E + i0) | g_{\nu} l \rangle)  .
\end{equation}

\noindent For more details of this construction we refer to
Refs. \cite{Haid84,Haid86a}. The chosen approximation energies $E_\mu$
for the interaction models considered in the present study are
summarized in Table \ref{seprep}.  These energies completely specify
the separable expansion (3.2) Note that, for reasons of convenience,
we have represented the (numerically given) form factors analytically
(cf. Eqs. (2.1) and (2.2) of Ref. \cite{Koike87a}), and have performed
the actual calculations with these expressions. The corresponding
parametrizations can be obtained from one of the authors (J.H.) on
request.

In Sec. V we are going to present three-nucleon bound state results,
based on the EST representation, with a varying rank in various
two-body partial waves.  We characterize these representations by
$(n_1n_2n_3...)$, where $n_1$, $n_2$, $n_3$, ... stand for the ranks
in the $^1s_0$, $^3s_1-{^3d_1}$, $^1p_1$, $^3p_0$, $^3p_1$, $^1d_2$,
$^3d_2$, and $^3p_2-{^3f_2}$ $NN$ partial waves. For a specific rank
$n_\mu$ in a certain partial wave the approximation energies $E_\mu$
can be read off from Table \ref{seprep}. They are given by the first
$n_\mu$ entries.  The notation in the results of the three-body
calculations is done in the same order of the partial waves, but the
ranks of the unused partial waves are left out.

\section{3N Bound--state calculation}
\label{secbound}

The triton bound state $| \Psi_{\rm t} \rangle $ is determined by the
eigenvalue equation

\begin{equation}
\label{eqeigen}
(E_{\rm t} - H)  \, |{\Psi_{\rm t}}\rangle = 0,
\end{equation}

\noindent
where the total Hamiltonian $H$ is given by $H = H_0 + V = H_0 +
\sum\limits_{\gamma = 1}^3 V_\gamma$.  Here we have used the
complementary notation $V_\gamma = V_{\alpha \beta}$ for the two-body
potentials, while $H_0$ denotes the free three-body Hamiltonian. When
introducing the corresponding resolvent $G_0(z) = (z - H_0)^{-1}$ or
the channel resolvents $G_\gamma(z) = (z - H_0 - V_\gamma)^{-1}$,
Eq. (\ref{eqeigen}) can be written in form of homogeneous integral
equations,

\begin{eqnarray}
\label{eqpsifull}
|{\Psi_{\rm t}}\rangle &=& G_{0}(E_{\rm t}) \,V \,|{\Psi_{\rm t}} \rangle =
G_{0}(E_{\rm t}) \sum_\gamma \, V_\gamma \,|{\Psi_{\rm t}} \rangle  \\
\label{eqpsifull2}
|{\Psi_{\rm t}}\rangle &=& G_{\gamma}(E_{\rm t}) \,\bar{V_{\gamma}} \,
|{\Psi_{\rm t}} \rangle = G_{\gamma}(E_{\rm t}) \sum_\beta (1 -
\delta_{ \gamma \beta})\, {V_{\beta}}\, |{\Psi_{\rm t}} \rangle,
\end{eqnarray}

\noindent
with $\bar{V_{\gamma}} = V - V_{\gamma}$ being the channel interaction
between particle $\gamma$ and the $(\alpha \beta$) subsystem.

\noindent
The latter equation can also be understood as a representation
of the bound state by the ``form-factors'' $|{F_\gamma}\rangle =
\bar{V_\gamma} \, |{\Psi_{\rm t}}\rangle$,

\begin{equation}
\label{eqpsiform}
 |{\Psi_{\rm t}}\rangle = G_{\gamma}(E_{\rm t}) \,|{F_\gamma}\rangle.
\end{equation}

\noindent
Multiplying this representation with $(1 - \delta_{\beta \gamma})
V_\gamma$, using the relation $V_\gamma \, G_\gamma = T_\gamma \, G_0$,
and summing over $\gamma$, we obtain for $|{F_\beta}\rangle$ the coupled
set of homogeneous integral equations

\begin{equation}
\label{eqform}
 |{F_\beta}\rangle = \sum_\gamma (1 - \delta_{\beta \gamma})\,
T_\gamma (E_{\rm t}) \,G_0(E_{\rm t})\, | {F_\gamma} \rangle .
\end{equation}

\noindent
Note that this relation may alternatively be derived by going to the
bound-state poles of the AGS equations, providing thus their homogeneous
version \cite{Alt67}. From (\ref{eqpsifull}) and (\ref{eqpsiform}) we
infer that the solutions of Eq.  (\ref{eqform}) provide $| \Psi_{\rm t}
\rangle$ according to

\begin{equation}
 |{\Psi_{\rm t}}\rangle = \sum_\gamma G_0(E_{\rm t}) \,
 T_\gamma(E_{\rm t}) \, G_0(E_{\rm t}) \, |{F_\gamma} \rangle =
 \sum_\gamma |{\psi_\gamma}\rangle.
\label{psi}
\end{equation}

\noindent
The $|{\psi_\gamma}\rangle$ are the standard Faddeev components, as seen
by using the definition of $|{F_\gamma}\rangle$ and the relation
(\ref{eqpsifull2}),

\begin{equation}
 |{\psi_\gamma}\rangle  =  G_0 \, T_\gamma \, G_0 \, |{F_\gamma} \rangle
= G_0 \, V_\gamma \, G_\gamma \, |{F_\gamma} \rangle
= G_0 \, V_\gamma G_\gamma \, \bar{V_\gamma} \,| \Psi_{\rm t} \rangle
= G_0 \, V_\gamma  | \Psi_{\rm t} \rangle .
\end{equation}

Equation (\ref{eqform}) will be treated numerically in momentum space,
employing a complete set of partial-waves states $| p \, q \,l\, b\,
\Gamma \,I \rangle$.  The label $b$ denotes the set $(\eta K L)$ of
quantum numbers, where $K$ and $L$ are the channel spin of the three
nucleons [with the coupling sequence $(j, \frac{1}{2}) K$] and the
relative angular momentum between the two-body subsystem and the third
particle, respectively.  $\Gamma$ is the total angular momentum
following from the coupling sequence $(K,L) \Gamma$, and $I$ is the
total isospin. These states satisfy the completeness relation

\begin{equation}
\label{complete}
1 = \sum_{b l \Gamma I} \int\limits_{0}^{\infty}
\int\limits_{0}^{\infty} dp \, p^2 \, dq \, q^2 \, | p \, q \,l\, b\,
\Gamma \,I \rangle \langle p \, q \,l\, b\, \Gamma \,I | .
\end{equation}

\noindent
The required antisymmetry under permutation of two particles in the
subsystem can be achieved by choosing only those states which satisfy
the condition $(-)^{l + s + t} = -1$ .  Table \ref{tabquantnum}
contains the quantum numbers of the corresponding channels taken into
account.

Inserting the separable {\sl T} matrix (\ref{eqtmatest}) for the EST
potentials and defining

\begin{equation}
 F^{\mu b}_\beta (q) = \sum_l \int\limits_{0}^{\infty} dp \, p^2 \,
g^{\eta}_{l \mu} (p) \,
 \langle p \, q \, l \, b\, \Gamma I | G_0 | F_\beta \rangle,
\end{equation}

\noindent
Eq.~(\ref{eqform}) goes over into
	
\begin{equation}
  F^{\mu b} (q) = \sum_{b'} \sum_{\nu \rho} \int\limits_{0}^{\infty} dq'
 \, q'^2 \, {\cal A V}^{b b'}_{\mu \nu} (q, q', E_{\rm t}) \,
 \Delta^{\eta'}_{\nu \rho} (E_{\rm t} - {\textstyle \frac{3}{4}} q'^2)
 \, F^{\rho b'} (q') ,
\label{form}
\end{equation}

\noindent
with

\begin{equation}
 {\cal A V}^{b b'}_{\mu \nu} (q, q', E_{\rm t}) = 2 \sum_{l l'}
\int\limits_{0}^{\infty} \int\limits_{0}^{\infty} dp \, p^2 \, dp' \,
p'^2 \, g^{\eta}_{l \mu}(p) \, \langle p \, q\, l\, b\, \Gamma \,I |
G_0(E_{\rm t}) | p' \,q'\, l'\, b'\, \Gamma \, I \rangle \,
g^{\eta'}_{l' \nu}(p')
\label{effpot}
\end{equation}

\noindent
being the so-called effective potential. The recoupling coefficients
entering this equation can be found in Ref. \cite{Janu93} (or in a
more compact form in \cite{Gloeck83} for another coupling sequence
which can easily be changed to the present one). In case of the {\sl
W}-matrix representation Eqs. (\ref{form}) and (\ref{effpot}) are of
similar form and, therefore, not given here.

After discretization Eq. (\ref{form}) can be treated as a linear
eigenvalue problem, where the energy is considered as a parameter
which is varied until the corresponding eigenvalue equals unity. The
eigenvalues can be found by using standard numerical algorithms. A
better approach is an iterative treatment, known as "power method"
\cite{Malt69}, which is justified due to the compactness of the kernel
of the integral equation employed.  It was found that this method is
much faster than standard eigenvalue algorithms and yields the same
accuracy.  For the direct solution of the two-dimensional Faddeev
equations we used a Lanczos-type algorithm \cite{Stadl91a,Saake92a}
that is even more efficient.

For the integration in Eq. (\ref{form}) a standard Gauss-Legendre mesh
was chosen. The angular integration in the effective potential was
done with 16 grid points. Table \ref{meshtab} contains results for the
Paris (EST) potential for different numbers of partial wave and an
increasing number of mesh points for the $q$ integration.  In all
cases 36 grid points were sufficient to get the binding energy up to 5
significant figures. For all further calculations we have used 40 mesh
points to get wave functions of high accuracy. In the calculations we
also included $q = 0$ to avoid extrapolations for small momenta in
further applications. The same was done for both variables in the
calculation of the wave function.  The binding energies obtained in
EST and {\sl W}-matrix approximation for different potentials are
given in Tables~\ref{bindparis}--\ref{bindbonnb}, compared with
results from a two-dimensional treatment \cite{Gloeck83} of the
Faddeev equations.

The whole wave function can now be calculated by either using Eq.
(\ref{psi}), or by applying the permutation operator $P$ on one Faddeev
component \cite{Gloeck83}

\begin{equation}
\label{eqperm}
| \Psi_{\rm t} \rangle = (1 + P) \, | \psi_1 \rangle,
\end{equation}

\noindent
where $P$ represents the sum of all cyclical and anticyclical
permutations of the nucleons.  From the practical point of view the
latter method has to be preferred.  Once $|\psi_1\rangle$ is computed
from $| \psi_1\rangle = G_0(E_{\rm t})\,T_1(E_{\rm t})\,G_0(E_{\rm t})
|F_1 \rangle$ the calculation of the full wave wave function via
Eq. (\ref{eqperm}) is independent of the rank of the separable
approximation, which considerably reduces the computing time.

The wave function is normalized according to

\begin{equation}
 \langle\Psi_{\rm t} | \Psi_{\rm t} \rangle = \sum_\gamma
\langle\psi_\gamma | \Psi_{\rm t} \rangle
= 3 \,  \langle\psi_1 | \Psi_{\rm t} \rangle = 1 .
\end{equation}

\noindent
It should be noted that, inserting the completeness relation
(\ref{complete}) into $\langle\Psi_{\rm t} | \Psi_{\rm t} \rangle$,
one has to deal with an infinite number of states due to the resulting
recouplings.  In contrast, when inserting the completeness relation in
$\langle\psi_1 | \Psi_{\rm t} \rangle$, one has to deal with a finite
number of partial waves corresponding to the ones in $\langle
\psi_1|$.  Plots of the wave functions for the Paris (EST) and
Bonn~{\sl A} (EST) potentials are shown in Figures 1 - 6. The figures
for the Bonn~{\sl B} (EST) potential are not distinguishable by eye
from the ones for the Bonn~{\sl A} (EST) potential, and are therefore
not shown.

\subsection{Properties of the wave function}

It is common to investigate the properties of the wave function in the
$LS$-coupling scheme (for simplicity we skip the dependence on the
isospins in the notation, since they are not recoupled)

\begin{equation}
|p \, q \, ((l L) {\cal L} (s S) {\cal S} ) \Gamma M_\Gamma \rangle .
\end{equation}

\noindent
In this scheme first the two orbital angular momenta and the two spins
are coupled separately. The total orbital angular momentum is then
coupled with the total spin to the total angular momentum of the
three-body system.  The total angular momentum of the triton is
$\Gamma = 1/2$, the total spin ${\cal S}$ of three particles can be
${\cal S } = 1/2$ or ${\cal S } = 3/2$.  The total orbital angular
momentum is, therefore, restricted to ${\cal L} = 0,1,2$.

The transformation from the channel spin into the $LS$ coupling scheme
is given by

\begin{eqnarray}
\label{transls}
& & \langle{((l L) {\cal L}, (s S) {\cal S}) \Gamma M_\Gamma }|
\nonumber  \\
& = & \sum_{j K} (-)^{l + s + L + S + {\cal L} + {\cal S } + 1}
\hat j \hat {\cal L} \hat{\cal S} \hat {K} \,
 \left\{ \!\!\begin{array}{ccc} {l}& {{\cal S}} & {K} \\
     {S}&{j}&{s}  \end{array} \!\!\right\} \,
 \left\{ \!\!\begin{array}{ccc} {l}&{{\cal S}}&{K}\\
    {\Gamma}&{L}&{{\cal L}} \end{array} \!\!\right\} \,
\langle{(((l s) j S) K L) \Gamma M_\Gamma}| ,
\end{eqnarray}

\noindent
with the abbreviation $\hat j = \sqrt{2 j + 1}$ used only in this
equation.  It should be noted that for this transformation the wave
function (\ref{eqperm}) has to be projected on all states that give a
contribution due to Eq. (\ref{transls}).  Otherwise the normalization
constant of the wave function is changed. It is not sufficient to use
only those channels used in the calculation of the Faddeev component.
Here again the recoupling of channels, as in the calculation of the
normalization constant, plays a role.

In the $LS$ coupling scheme the wave function can be classified
according to the contributions of the states belonging to ${\cal L} =
0,1,2$.

\begin{eqnarray}
1 &=& 3 \;\langle\psi_1 | \Psi_{\rm t} \rangle \nonumber \\ & = &
3 \sum_{\cal L} \sum_{\cal S} \sum_{l L s S }
\int\limits_{0}^{\infty}\int\limits_{0}^{\infty} dp \, p^2 \, dq \,q^2
\, \langle\psi_1 | p \, q \, ((l L) {\cal L} (s S) {\cal S} ) \Gamma
\rangle \; \langle p \, q \, ((l L) {\cal L} (S s) {\cal S} ) \Gamma|
\Psi_{\rm t} \rangle \nonumber \\ & = & \sum_{\cal L} {\rm P}({\cal
L}) = {\rm P(S)} + {\rm P(S')} + {\rm P(P)} + {\rm P(D)}.
\end{eqnarray}

\noindent
The contributions to the normalization constant for a certain total
angular momentum are denoted by P(${\cal L}$). In case of ${\cal L} = 0$
also the symmetric and mixed symmetric spatial contributions P(S) and
P(S'), are extracted. The antisymmetric part P(S'') of the
wave function is negligible and, therefore, has been omitted. The main
contribution to the mixed symmetric part stems from the difference
between the $^1s_0$ and the $^3s_1$ interaction.

\section{Results and Conclusions}

As a test of accuracy of the EST and {\sl W}-matrix approaches we
compare the triton binding energies and wave functions obtained in
this way with the results of a two-dimensional treatment of the
Faddeev equations. In Tables \ref{bindparis}-\ref{bindbonnb} our
binding energies obtained for the Paris, Bonn {\sl A}, and Bonn {\sl
B} potentials are given for different combinations of partial waves.
For all three potentials considered here, the EST results converge
with increasing rank and agree within 0.2\% with the two-dimensional
results.  In case of the {\sl W} matrix, the free parameter $k$ was
chosen differently in each partial wave according to the criterion
employed in Ref.  \cite{Janu93}, which consists in providing a binding
energy close to the results obtained with other methods. This
optimization method is not fully satisfactory in the $j \leq 2$
calculations, due to ambiguities in fixing $k$ when including the
$^3p_2$-$^3f_2$ partial wave. The agreement of the {\sl W}-matrix
calculations with the two-dimensional results is less good than in the EST
case. There are differences between 0 $\leq$ 4\%.

Our five-channel, i.e., $j \leq 1^+$, calculations for the Paris (EST)
potential are also in perfect agreement with the results by Parke et
al. \cite{Park91}.  We can, moreover, compare with the results by
Friar et al.  \cite{Fria88c} generated in coordinate space, and find
again good agreement. For $j \leq 2$ and the Paris potential our results
are exactly the same as those of Ref. \cite{Nemoto98a}. The binding
energies for Bonn {\sl A} and $j \leq 2$ differ only slightly from those
by Fonseca and Lehman \cite{Fons95a}.  

For the Paris (EST), Bonn {\sl A} (EST), Bonn {\sl B} (EST) potentials
the components of the three-body wave function, defined in the
previous section, differ from the two-dimensional results by 0-2.5\%,
0-0.05\%, and 0-3\% respectively.  It should be emphasized, however,
that only P(P) shows the large deviation of 3\% quoted, while all
other components are in much better agreement.

The {\sl W}-matrix results for the components of the wave function
differ by 0-12\%, 0-10\%, and 0-10\% in case of the Paris, Bonn {\sl A},
and Bonn {\sl B} potentials, respectively.  Here the largest deviations
are found both in the P(P) and in the P(S') components.

Another test is given by the norm squared of the differences of the
triton wave functions obtained in the separable and the
two-dimensional treatments, $ \Delta N = || \Psi_{\rm {sep}} -
\Psi_{\rm 2d}||^2$. For the EST potentials $\Delta N$ is of the order
$10^{-6}$, while for the {\sl W}-matrix representation it is of the
order $10^{-5}$. Also in this respect the EST method, hence, leads to
better results.

Thus, we have demonstrated the high quality of the EST expansion
method. For $j \leq 1^+$ and the Paris (EST) interaction this has been
done already in \cite{Park91}, and for the Argonne potential (with a
modified expansion scheme) in \cite{Koike97a}. Here we have extended
these former investigations up to $j \leq 2$, using moreover two
versions of the Bonn potential. The accuracy achieved within the {\sl
W}-matrix approach is less satisfactory, namely of the order
2-10\%. But it should be recalled that this treatment is based on a
rather simple rank-one approximation only.

\begin{acknowledgments}
The work of W.~S. and W.~Sch., and that of A. N. was supported by the
Deutsche Forschungsgemeinschaft under Grant Nos. Sa 327/23-1 and 
GL-8727-1, respectively.  Part of this work has been done under the
auspices of the U.~S.  Department of Energy under contract
No. DE-FG02-93ER40756 with Ohio University. The numerical calculations
were partly performed on the Cray T3E of the
H\"ochstleistungsrechenzentrum in J\"ulich, Germany.
\end{acknowledgments}

\begin{table}[ht] \begin{center} \begin{tabular}{lccccccc}
\multicolumn{8}{c} {Paris potential} \\ \hline
partial wave &\multicolumn{7}{c} {$(E_\mu,l_\mu)$} \\ \hline
$^1s_0$ & 0 & 100 & 500 & -100 & -200 & &  \\
$^3s_1-{^3d_1}$ & $\epsilon_d$ &
(100,0) & (125,2) & (425,2) & (-50,0) & (-50,2) & \\
$^1p_1$,$^3p_1$ & 10 & -50 & 150 & 300 & -150 & &  \\
$^3p_0$ & 10 & -50 & 150 & 350 & -150 & &  \\
$^3p_2-{^3f_2}$ & (10,1)  &
(40,3) & (75,1) & (75,3) & (175,1) & (175,3) & (300,1) \\
$^1d_2$,$^3d_2$ & 10 & -50 & 150 & 300 & -150 & &  \\
 \\ \hline \hline
\multicolumn{8}{c} {Bonn {\sl A} and {\sl B} potentials} \\ \hline
partial wave &\multicolumn{7}{c} {$(E_\mu,l_\mu)$} \\ \hline
$^1s_0$ & 0 & 100 & 300 & -100 & -50 & &  \\
$^3s_1-{^3d_1}$ & $\epsilon_d$ &
(50,2) & (100,0) & (300,2) & (-50,0) & (-50,2) & \\
$^1p_1$,$^3p_0$,$^3p_1$ & 10 & -50 & 150 & 300 & -150 & &  \\
$^3p_2-{^3f_2}$ & (10,1)  &
(10,3) & (75,1) & (75,3) & (150,1) & (150,3) & (200,1) \\
$^1d_2$,$^3d_2$ & 10 & -50 & 150 & 300 & -150 & &  \\
 \\ \end{tabular} \end{center}
\caption{Approximation energies $E_\mu$ used in the EST
representations of the Paris, Bonn {\sl A} and Bonn {\sl B}
potentials.  $\epsilon_d$ refers to the deuteron binding energy.
$E_\mu$ are lab energies in MeV.  In case of coupled partial waves,
the boundary condition chosen for the angular momentum $l_\mu$ of the
initial state (cf. Ref. \protect\cite{Haid84}) is also specified.}
\label{seprep} \end{table}

\begin{table}[ht]
\begin{center}
\begin{tabular}{cccccccc}
 Channel & Subsystem & $l$ & $s$ & $j^\pi$ & $\tau$ & $K$ & $L$ \\
\hline
 1 & $^1s_0$ & $0$ & $0$ & $0^+$ & $1$ & $1/2$ & $0$ \\
 2 & $^3s_1$ & $0$ & $1$ & $1^+$ & $0$ & $1/2$ & $0$ \\
 3 & $^3s_1$ & $0$ & $1$ & $1^+$ & $0$ & $3/2$ & $2$ \\
 4 & $^3d_1$ & $2$ & $1$ & $1^+$ & $0$ & $1/2$ & $0$ \\
 5 & $^3d_1$ & $2$ & $1$ & $1^+$ & $0$ & $3/2$ & $2$ \\
 6 & $^3p_0$ & $1$ & $1$ & $0^-$ & $1$ & $1/2$ & $1$ \\
 7 & $^1p_1$ & $1$ & $0$ & $1^-$ & $0$ & $1/2$ & $1$ \\
 8 & $^1p_1$ & $1$ & $0$ & $1^-$ & $0$ & $3/2$ & $1$ \\
 9 & $^3p_1$ & $1$ & $1$ & $1^-$ & $1$ & $1/2$ & $1$ \\
10 & $^3p_1$ & $1$ & $1$ & $1^-$ & $1$ & $3/2$ & $1$ \\
11 & $^1d_2$ & $2$ & $0$ & $2^+$ & $1$ & $3/2$ & $2$ \\
12 & $^1d_2$ & $2$ & $0$ & $2^+$ & $1$ & $5/2$ & $2$ \\
13 & $^3d_2$ & $2$ & $1$ & $2^+$ & $0$ & $3/2$ & $2$ \\
14 & $^3d_2$ & $2$ & $1$ & $2^+$ & $0$ & $5/2$ & $2$ \\
15 & $^3p_2$ & $1$ & $1$ & $2^-$ & $1$ & $3/2$ & $1$ \\
16 & $^3p_2$ & $1$ & $1$ & $2^-$ & $1$ & $5/2$ & $3$ \\
17 & $^3f_2$ & $3$ & $1$ & $2^-$ & $1$ & $3/2$ & $1$ \\
18 & $^3f_2$ & $3$ & $1$ & $2^-$ & $1$ & $5/2$ & $3$ \\
\end{tabular} \end{center}
\caption{Quantum numbers of the three-body channels.}
\label{tabquantnum}
\end{table}

\begin{table}[ht]
\begin{center}
\begin{tabular}{r|ccc}
\# Meshpoints  & $j \leq 1^+ $ & $j \leq 1$ & $j \leq 2$ \\
\hline
 6     & -8.3318 & -8.1383 & -9.0551 \\
 12    & -7.3571 & -7.1376 & -7.4330 \\
 24    & -7.3150 & -7.0913 & -7.3688 \\
 36    & -7.3156 & -7.0919 & -7.3688 \\
 40    & -7.3156 & -7.0919 & -7.3688 \\
\end{tabular} \end{center}
\caption{Triton binding energies (in MeV) with the Paris (EST)
potential (56555557). The notation (56...) specifies the employed
separable representation as explained in Sec. \ref{secestmethod}. }
\label{meshtab}
\end{table}

\begin{table}[ht]
\begin{center}
\begin{tabular}{llccccc}
 & & $E_{\rm t}$(MeV) & P(S) & P(S') & P(P) & P(D) \\
\hline
 $j \leq 1^+$
 & Paris (EST) (11) & -7.451 & 90.63 & 1.636 & 0.042 & 7.692 \\
 & Paris (EST) (34) & -7.266 & 89.88 & 1.652 & 0.065 & 8.402 \\
 & Paris (EST) (56) & -7.316 & 89.90 & 1.634 & 0.064 & 8.401 \\
 \hline
 & Paris ({\sl W}-matrix) & -7.300 & 90.22 & 1.450 & 0.064 & 8.265 \\
\hline
 & Paris                  & -7.297 & 89.88 & 1.625 & 0.066 & 8.428 \\
\hline
 & Paris-r                & -7.310 & 89.88 & 1.623 & 0.066 & 8.428 \\
\hline \hline
 $j \leq 2^+$
 & Paris (EST) (1111) & -7.464 & 90.62 & 1.636 & 0.042 & 7.704 \\
 & Paris (EST) (3444) & -7.375 & 89.87 & 1.618 & 0.066 & 8.447 \\
 & Paris (EST) (5644) & -7.424 & 89.89 & 1.601 & 0.066 & 8.444 \\
 & Paris (EST) (5655) & -7.426 & 89.89 & 1.600 & 0.066 & 8.446 \\
\hline
 & Paris ({\sl W}-matrix) & -7.343 & 90.21 & 1.436 & 0.064 & 8.288 \\
\hline
 & Paris                  & -7.408 & 89.87 & 1.591 & 0.068 & 8.474 \\
\hline \hline
 $j \leq 1$
 & Paris (EST) (11111) & -7.464 & 90.83 & 1.468 & 0.044 & 7.658 \\
 & Paris (EST) (34333) & -7.074 & 90.28 & 1.492 & 0.064 & 8.167 \\
 & Paris (EST) (56444) & -7.093 & 90.30 & 1.488 & 0.062 & 8.154 \\
 & Paris (EST) (56555) & -7.092 & 90.30 & 1.488 & 0.062 & 8.153 \\
\hline
 & Paris ({\sl W}-matrix) & -7.150 & 90.51 & 1.301 & 0.066 & 8.123 \\
\hline
 & Paris                  & -7.103 & 90.28 & 1.468 & 0.063 & 8.193 \\
\hline \hline
 $j\leq 2$
 & Paris (EST) (11111111)& -7.549 & 90.61 & 1.459 & 0.047 & 7.879 \\
 & Paris (EST) (56555557)& -7.369 & 90.14 & 1.420 & 0.063 & 8.379 \\
\hline
 & Paris ({\sl W}-matrix) & -7.088 & 90.54 & 1.366 & 0.062 & 8.034 \\
\hline
 & Paris                  & -7.378 & 90.11 & 1.403 & 0.064 & 8.418 \\
\end{tabular}
\end{center}
\caption{Triton wave function components for the Paris potential. The
notation (56...) etc. specifies the employed separable (EST)
representation as explained in Sec. \ref{secestmethod}. The result
for Paris-r is taken from Ref.  \protect\cite{Fria88c}.}
\label{bindparis} \end{table}

\begin{table}[ht]
\begin{center}
 \begin{tabular}{llccccc}
 & & $E_{\rm
t}$(MeV) & P(S) & P(S') & P(P) & P(D) \\
 \hline
 $j \leq 1^+$
 & Bonn {\sl A} (EST) (11) & -8.350 & 92.75 & 1.415 & 0.028 & 5.811
\\
 & Bonn {\sl A} (EST) (44) & -8.347 & 92.35 & 1.427 & 0.034 & 6.188 \\
 & Bonn {\sl A} (EST) (56) & -8.380 & 92.31 & 1.432 & 0.035 & 6.220 \\
\hline
 & Bonn {\sl A} ({\sl W}-matrix) & -8.371 & 92.01 & 1.385 & 0.041 & 6.565 \\
\hline
 & Bonn {\sl A}                  & -8.378 & 92.32 & 1.426 & 0.035 & 6.217 \\
\hline \hline
 $j \leq 2^+$
 & Bonn {\sl A} (EST) (1111) & -8.360 & 92.74 & 1.415 & 0.027 & 5.820 \\
 & Bonn {\sl A} (EST) (4444) & -8.411 & 92.35 & 1.411 & 0.034 & 6.204 \\
 & Bonn {\sl A} (EST) (5644) & -8.444 & 92.31 & 1.415 & 0.034 & 6.236 \\
\hline
 & Bonn {\sl A} ({\sl W}-matrix) & -8.399 & 92.00 & 1.380 & 0.039 & 6.578 \\
\hline
 & Bonn {\sl A}                  & -8.443 & 92.32 & 1.411 & 0.035 & 6.235 \\
\hline \hline
 $j \leq 1$
 & Bonn {\sl A} (EST) (11111) & -8.298 & 92.95 & 1.248 & 0.030 & 5.772 \\
 & Bonn {\sl A} (EST) (44444) & -8.083 & 92.72 & 1.252 & 0.037 & 5.995 \\
 & Bonn {\sl A} (EST) (56444) & -8.115 & 92.68 & 1.254 & 0.037 & 6.027 \\
\hline
 & Bonn {\sl A} ({\sl W}-matrix) & -8.160 & 92.36 & 1.209 & 0.043 & 6.391 \\
\hline
 & Bonn {\sl A}                  & -8.127 & 92.69 & 1.248 & 0.037 & 6.029 \\
\hline \hline
 $j \leq 2$
 & Bonn {\sl A} (EST) (11111111) & -8.395 & 92.81 & 1.264 & 0.031 & 5.895 \\
 & Bonn {\sl A} (EST) (56444445) & -8.285 & 92.59 & 1.236 & 0.037 & 6.135 \\
\hline
 & Bonn {\sl A}                  & -8.295 & 92.59 & 1.231 & 0.037 & 6.138 \\
\end{tabular}
\end{center}
\caption{Triton wave function components for the Bonn {\sl A} potential.
(56...) etc. specifies the employed separable (EST) representation as
explained in Sec. \ref{secestmethod}. }
\label{bindbonna}  \end{table}

\begin{table}[ht]
\begin{center}
\begin{tabular}{llccccc} & & $E_{\rm t}$(MeV) & P(S) & P(S') & P(P) & P(D) \\
\hline $j \leq 1^+$
 & Bonn {\sl B} (EST) (11) & -8.209 & 91.95 & 1.361 & 0.035 & 6.659 \\
 & Bonn {\sl B} (EST) (44) & -8.137 & 91.36 & 1.369 & 0.047 & 7.224 \\
 & Bonn {\sl B} (EST) (56) & -8.170 & 91.34 & 1.373 & 0.048 & 7.236 \\
 \hline
 & Bonn {\sl B} ({\sl W}-matrix) & -8.161 & 91.21 & 1.286 & 0.053 & 7.448 \\
\hline
 & Bonn {\sl B}                  & -8.165 & 91.35 & 1.368 & 0.049 & 7.235 \\
 \hline \hline
 $j \leq 2^+$
 & Bonn {\sl B} (EST) (1111) & -8.219 & 91.94 & 1.360 & 0.034 & 6.668 \\
 & Bonn {\sl B} (EST) (4444) & -8.197 & 91.35 & 1.354 & 0.047 & 7.245 \\
 & Bonn {\sl B} (EST) (5644) & -8.230 & 91.34 & 1.357 & 0.048 & 7.256 \\
 \hline
 & Bonn {\sl B} ({\sl W}-matrix) & -8.190 & 91.21 & 1.286 & 0.053 & 7.448 \\
 \hline
 & Bonn {\sl B}                  & -8.226 & 91.34 & 1.354 & 0.049 & 7.257 \\
\hline \hline
 $j \leq 1$
 & Bonn {\sl B} (EST) (11111) & -8.159 & 92.14 & 1.200 & 0.037 & 6.620 \\
 & Bonn {\sl B} (EST) (44444) & -7.855 & 91.75 & 1.216 & 0.048 & 6.985 \\
 & Bonn {\sl B} (EST) (56444) & -7.884 & 91.74 & 1.218 & 0.048 & 6.700 \\
\hline
 & Bonn {\sl B} ({\sl W}-matrix) & -7.926 & 91.56 & 1.143 & 0.054 & 7.247 \\
\hline
 & Bonn {\sl B}                  & -7.899 & 91.74 & 1.210 & 0.049 & 7.001 \\
\hline \hline
  $j \leq 2$
 & Bonn {\sl B} (EST) (11111111) & -8.282 & 91.96 & 1.208 & 0.039 & 6.791 \\
 & Bonn {\sl B} (EST) (56444445) & -8.088 & 91.61 & 1.189 & 0.048 & 7.149 \\
 \hline
 & Bonn {\sl B} ({\sl W}-matrix) & -7.919 & 91.56 & 1.143 & 0.052 & 7.249 \\
\hline
 & Bonn {\sl B}                  & -8.103 & 91.62 & 1.184 & 0.048 & 7.152 \\
\end{tabular}
\end{center}
\caption{Triton wave function components for the Bonn {\sl B} potential.
(56...) etc. specifies the employed separable (EST) representation as
explained in Sec. \ref{secestmethod}. }
\label{bindbonnb} \end{table}

\input{psfig}

\begin{figure}
\centerline{\hbox{
\psfig{figure=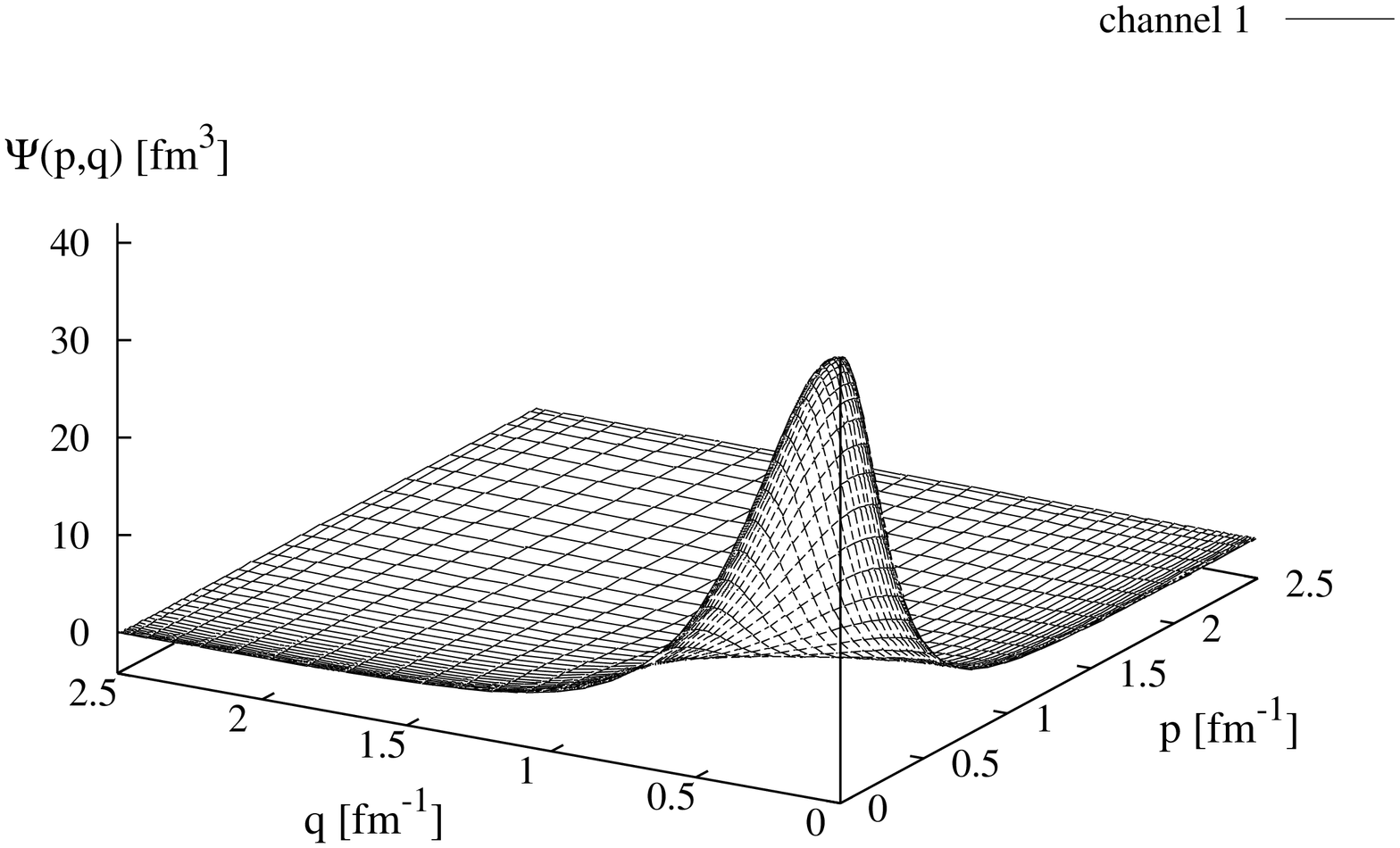,height=6.9cm,width=7.7cm,angle=-90}
\hspace{0.5cm}
\psfig{figure=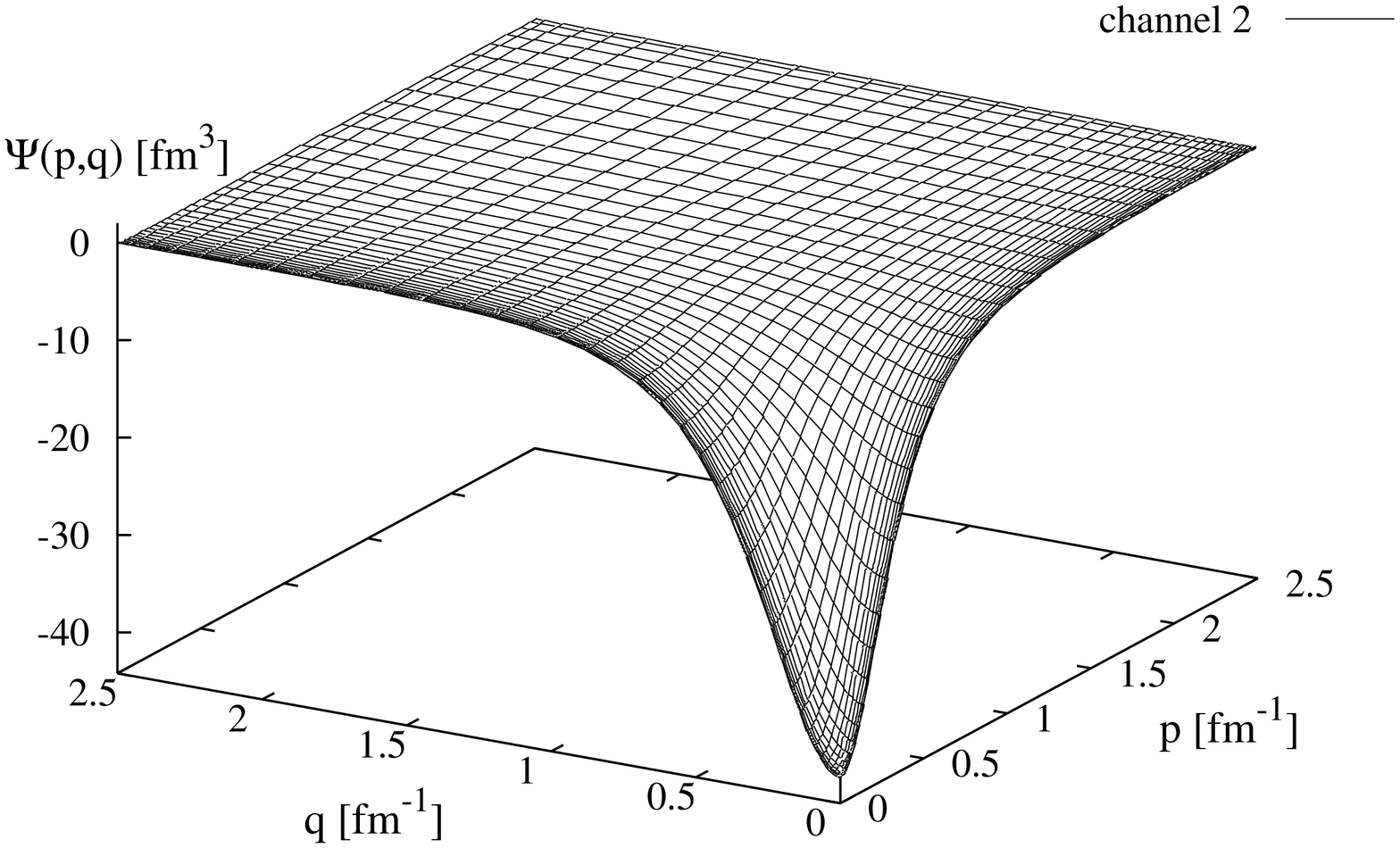,height=6.9cm,width=7.7cm,angle=-90}
}}
\centerline{\hbox{
\psfig{figure=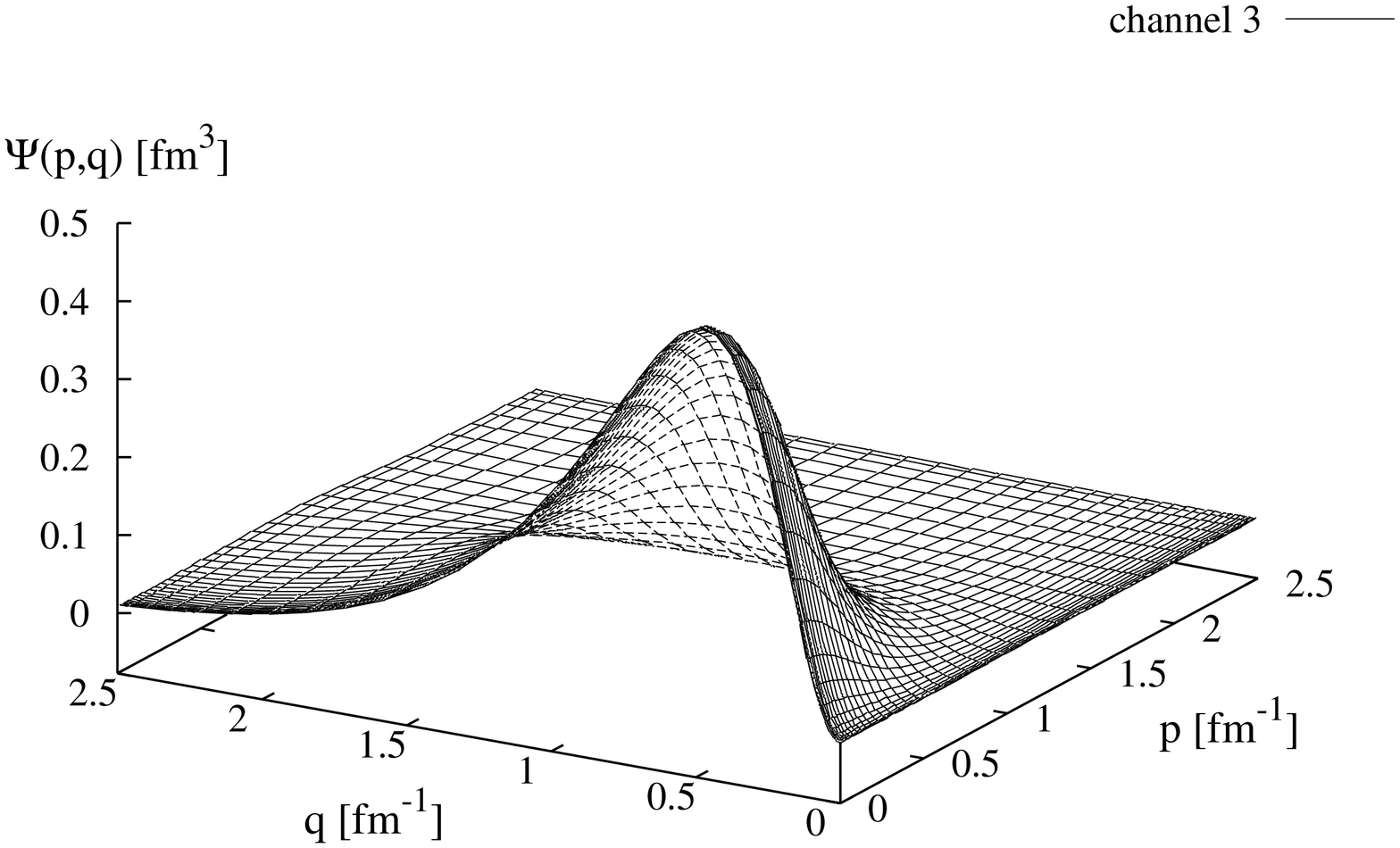,height=6.9cm,width=7.7cm,angle=-90}
\hspace{0.5cm}
\psfig{figure=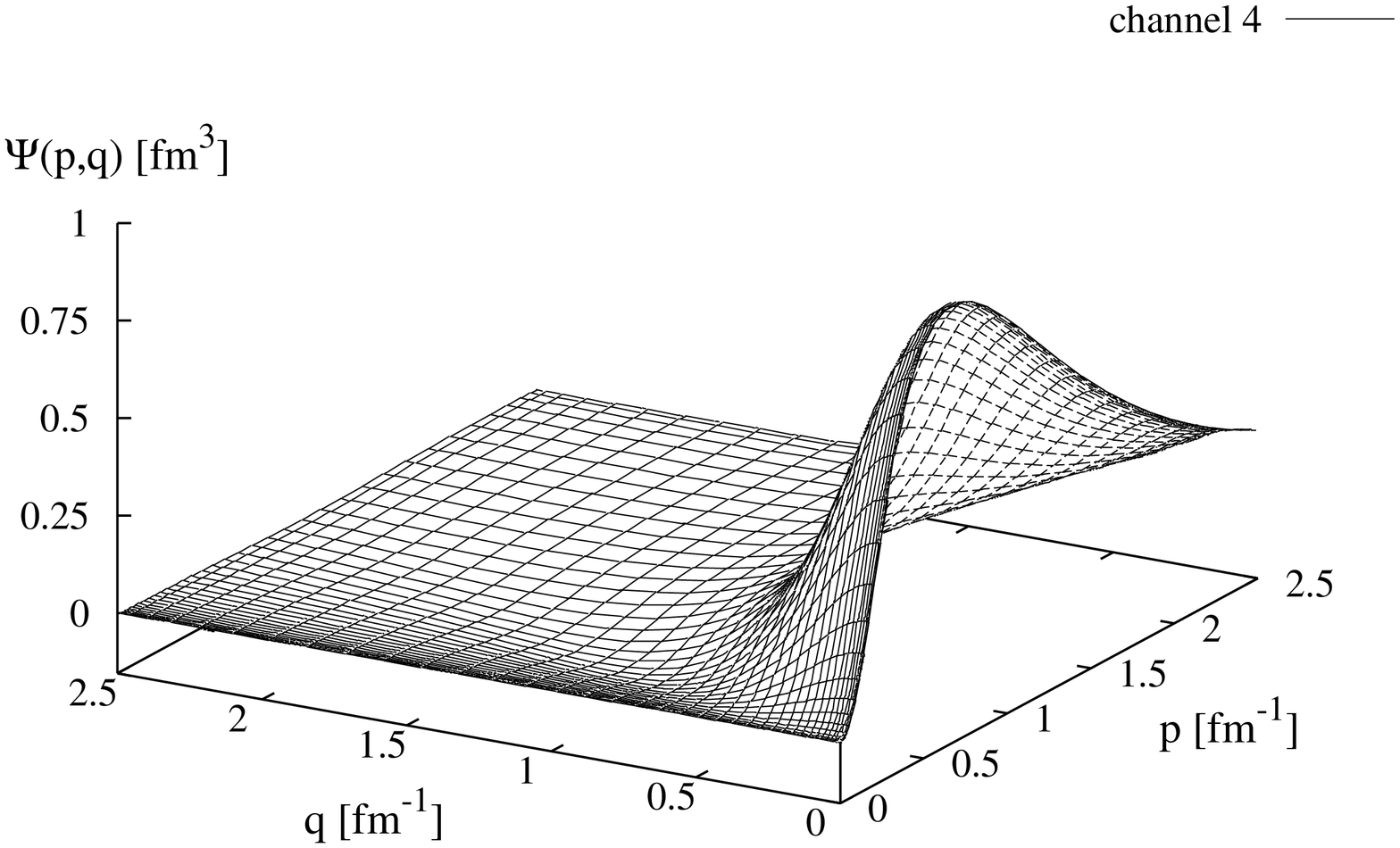,height=6.9cm,width=7.7cm,angle=-90}
}}
\centerline{\hbox{
\psfig{figure=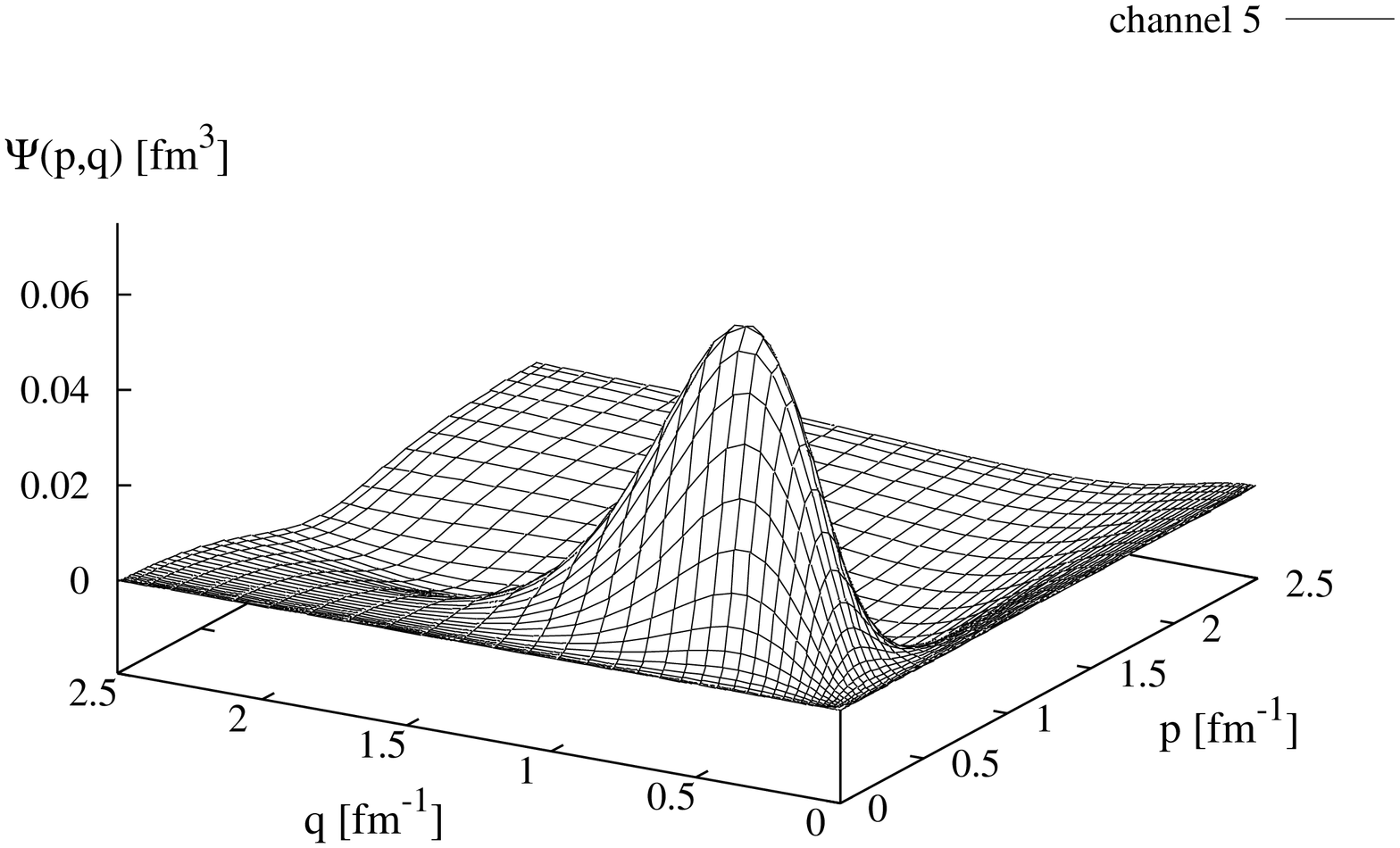,height=6.9cm,width=7.7cm,angle=-90}
\hspace{0.5cm}
\psfig{figure=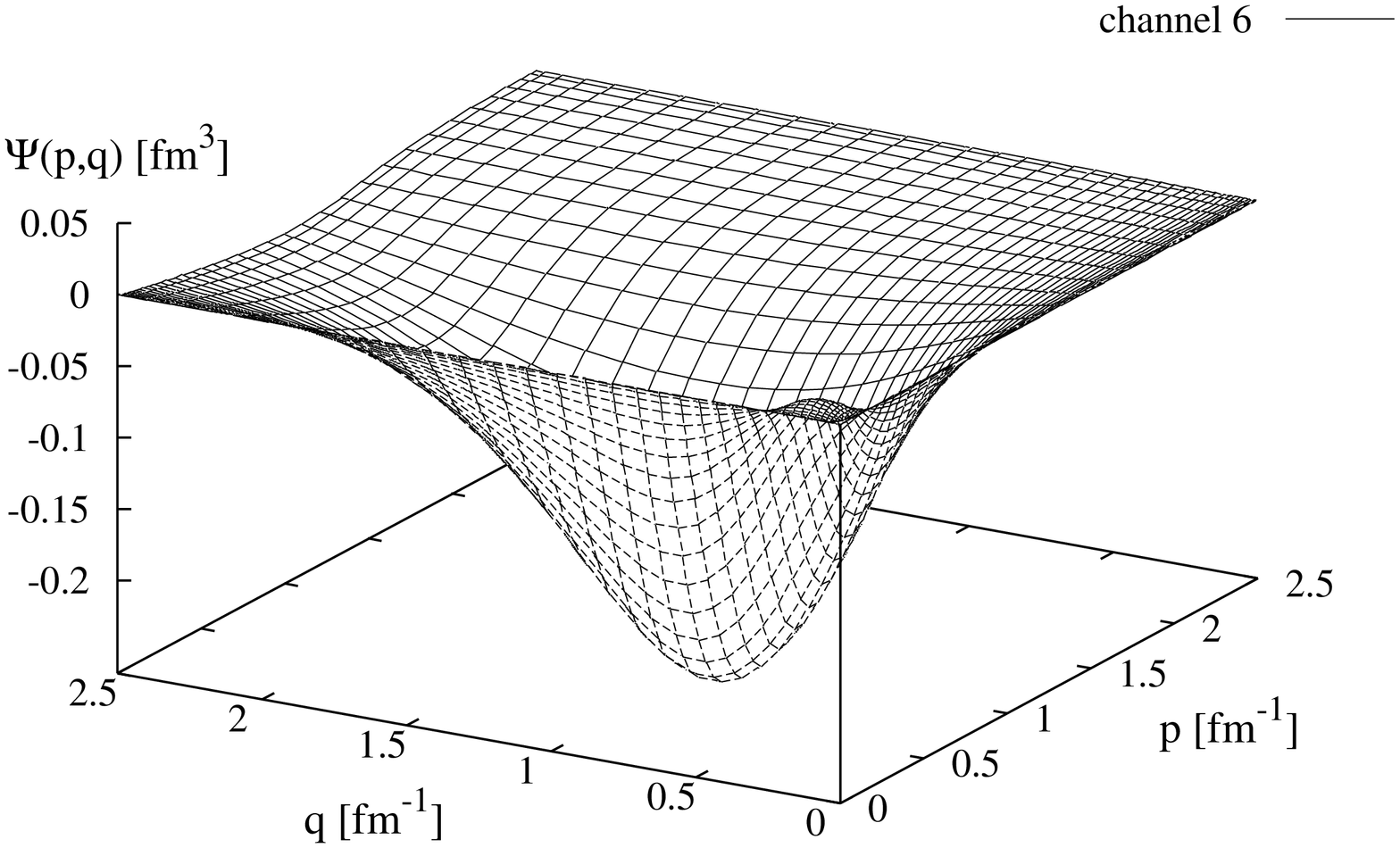,height=6.9cm,width=7.7cm,angle=-90}
}}
\caption{Graphs of the triton wave function obtained with the Paris
(EST) potential. The channels are defined in Table \ref{tabquantnum}.}
\end{figure}

\begin{figure}
\centerline{\hbox{
\psfig{figure=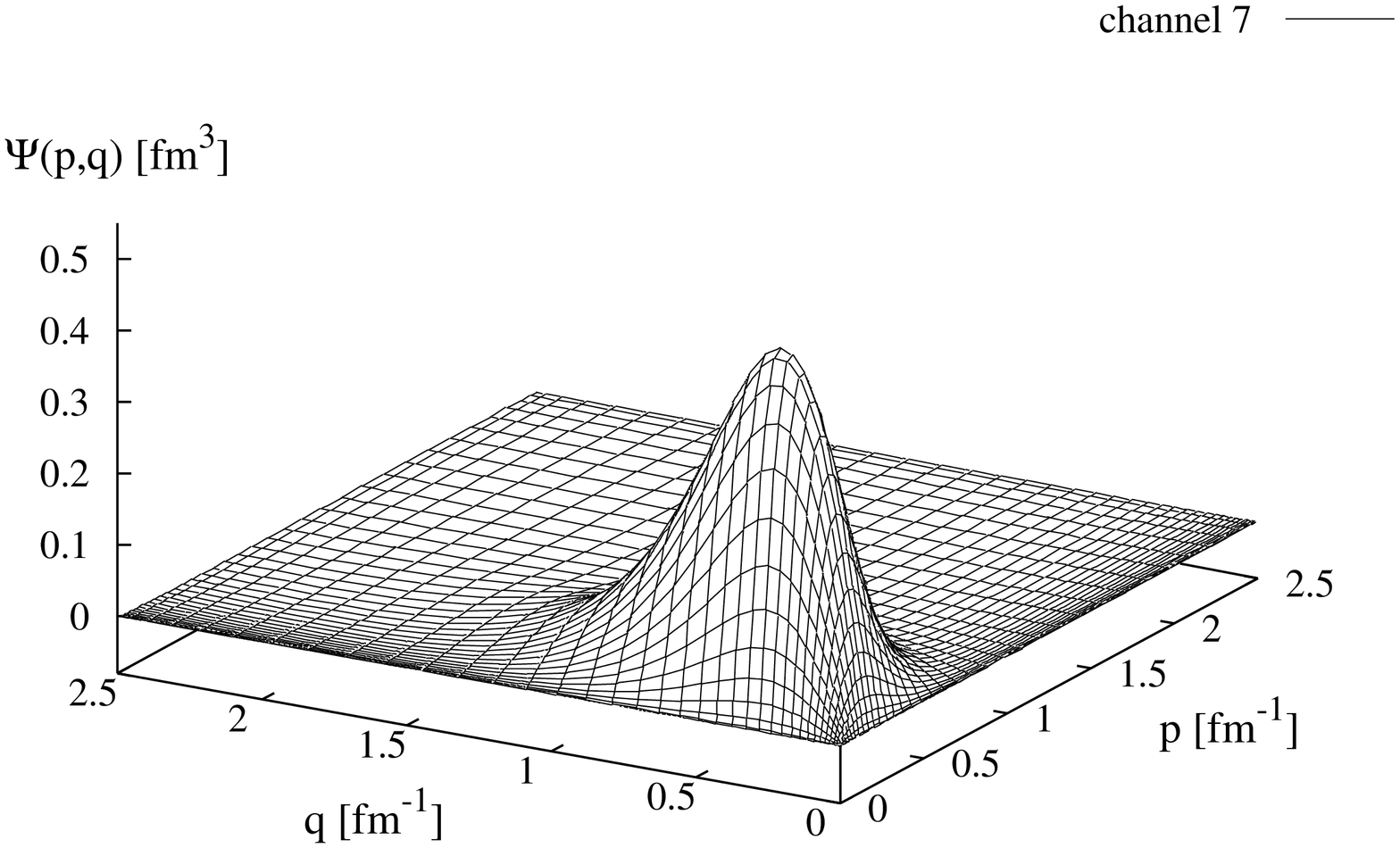,height=6.9cm,width=7.7cm,angle=-90}
\hspace{0.5cm}
\psfig{figure=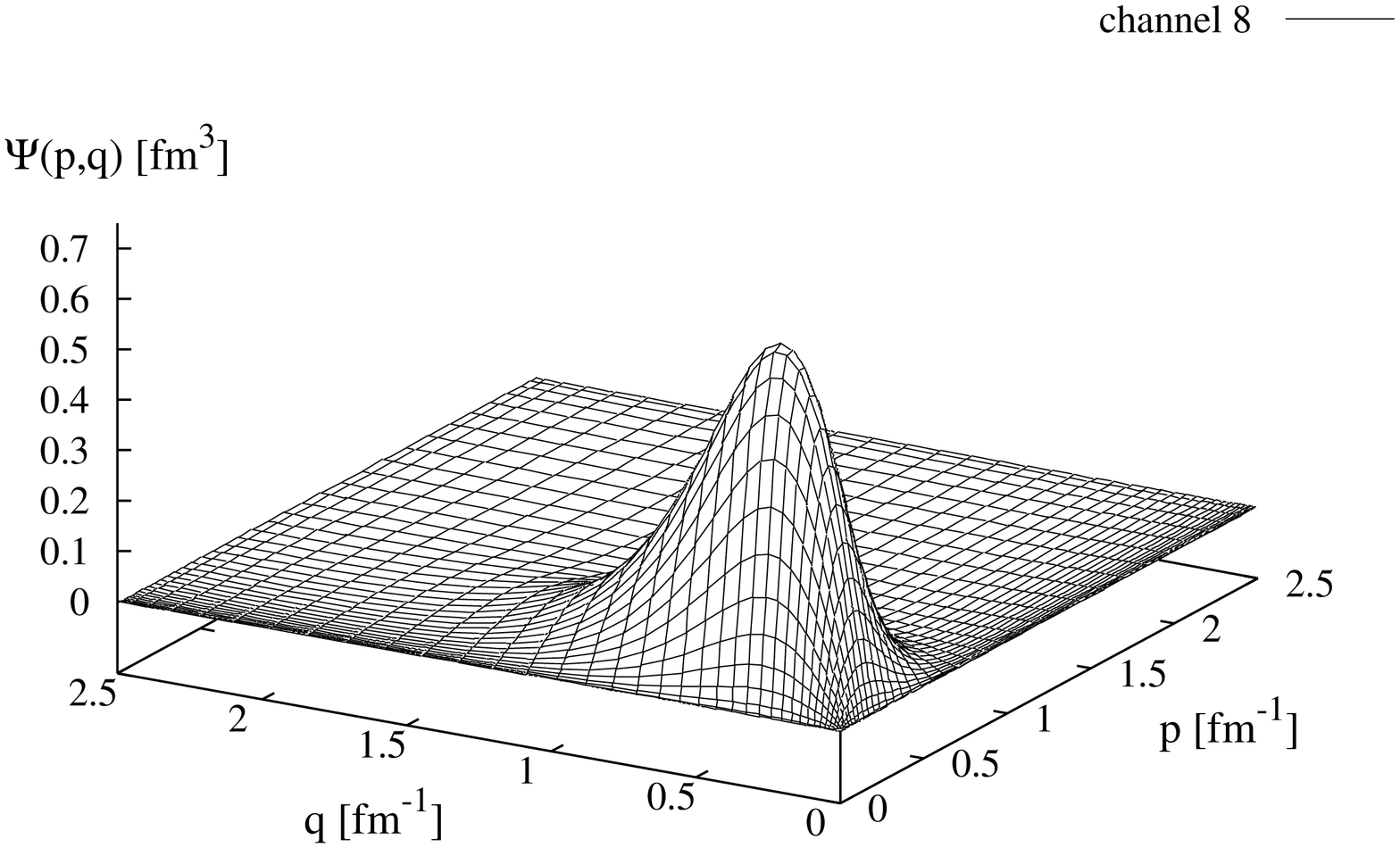,height=6.9cm,width=7.7cm,angle=-90}
}}
\centerline{\hbox{
\psfig{figure=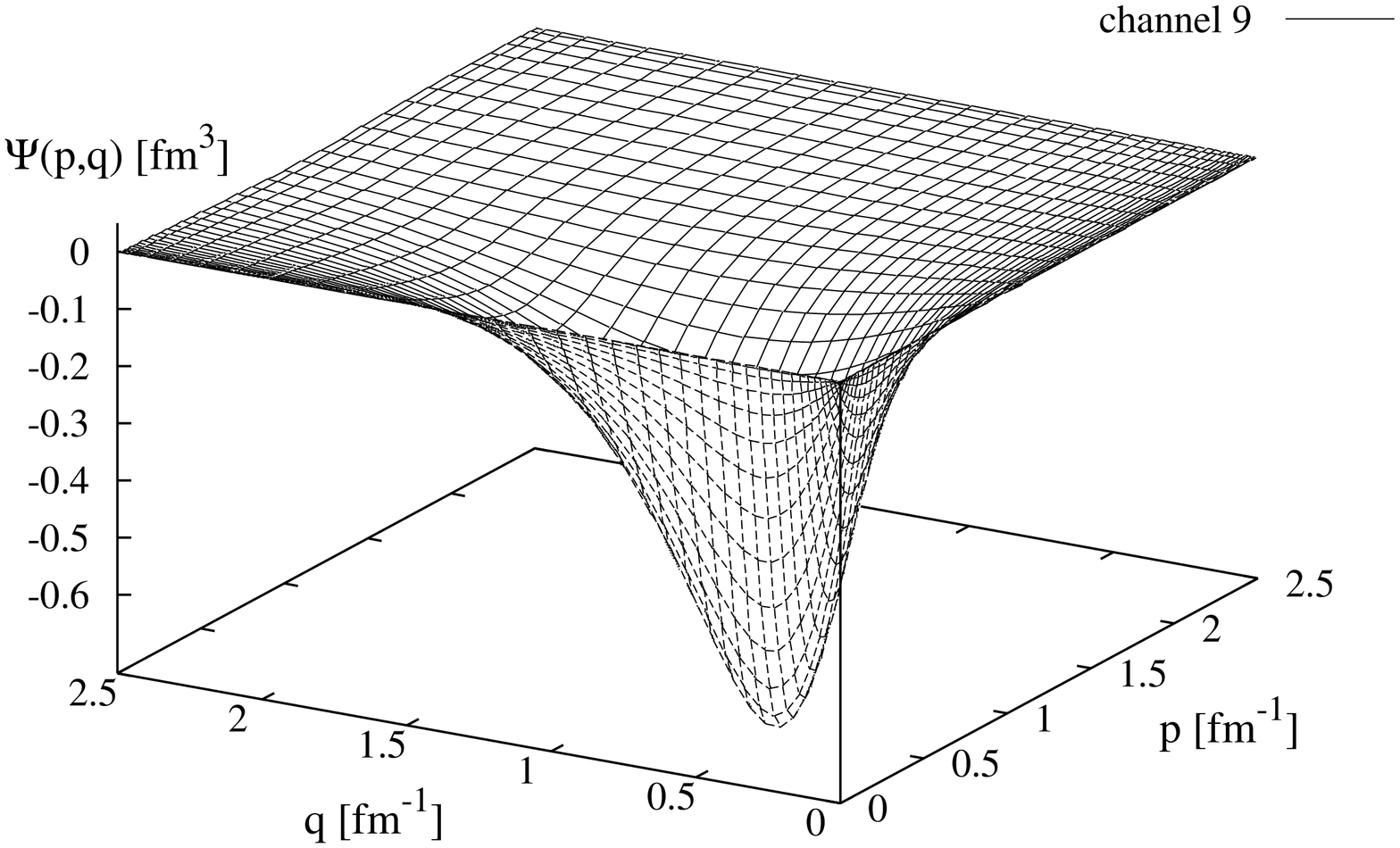,height=6.9cm,width=7.7cm,angle=-90}
\hspace{0.5cm}
\psfig{figure=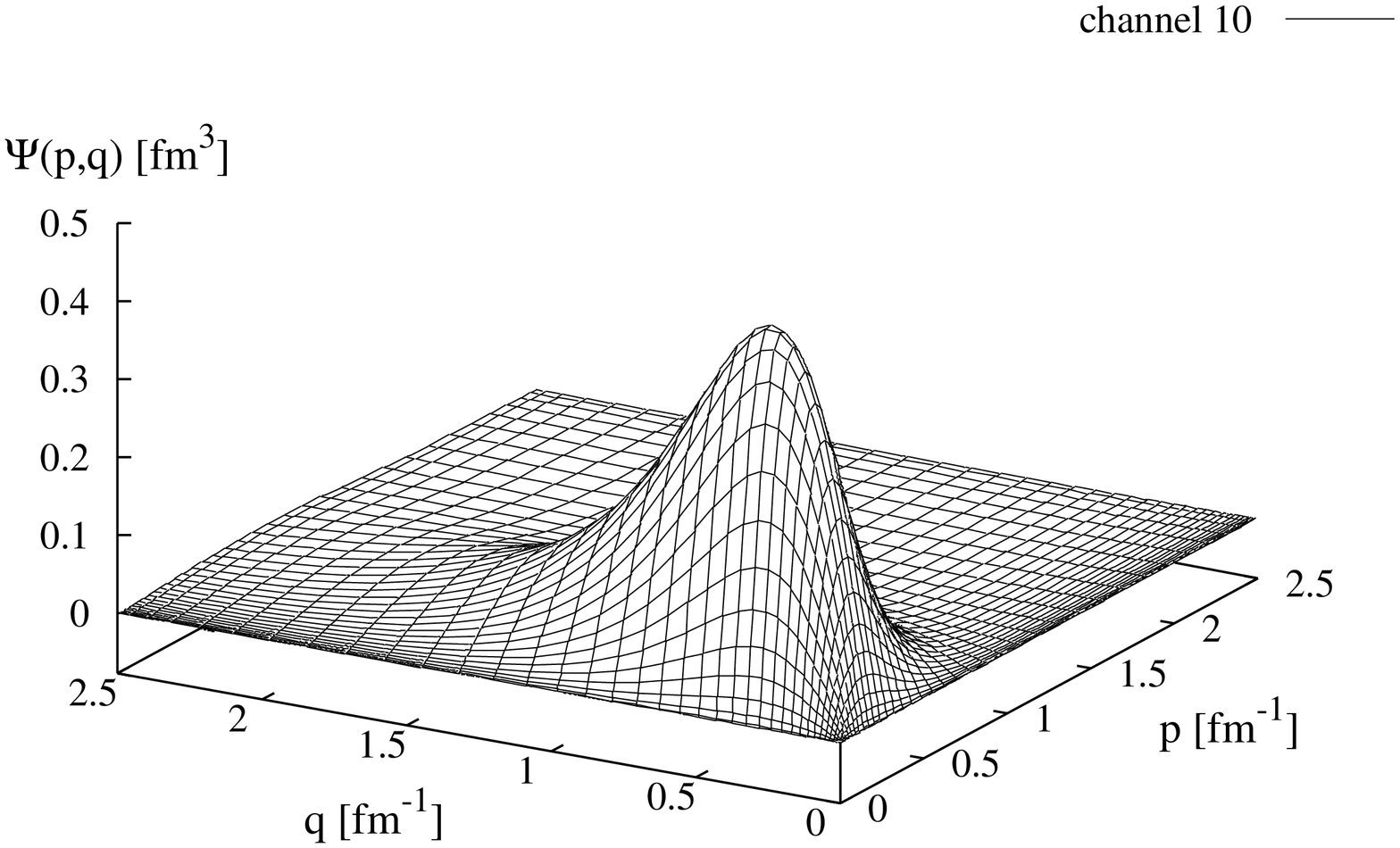,height=6.9cm,width=7.7cm,angle=-90}
}}
\centerline{\hbox{
\psfig{figure=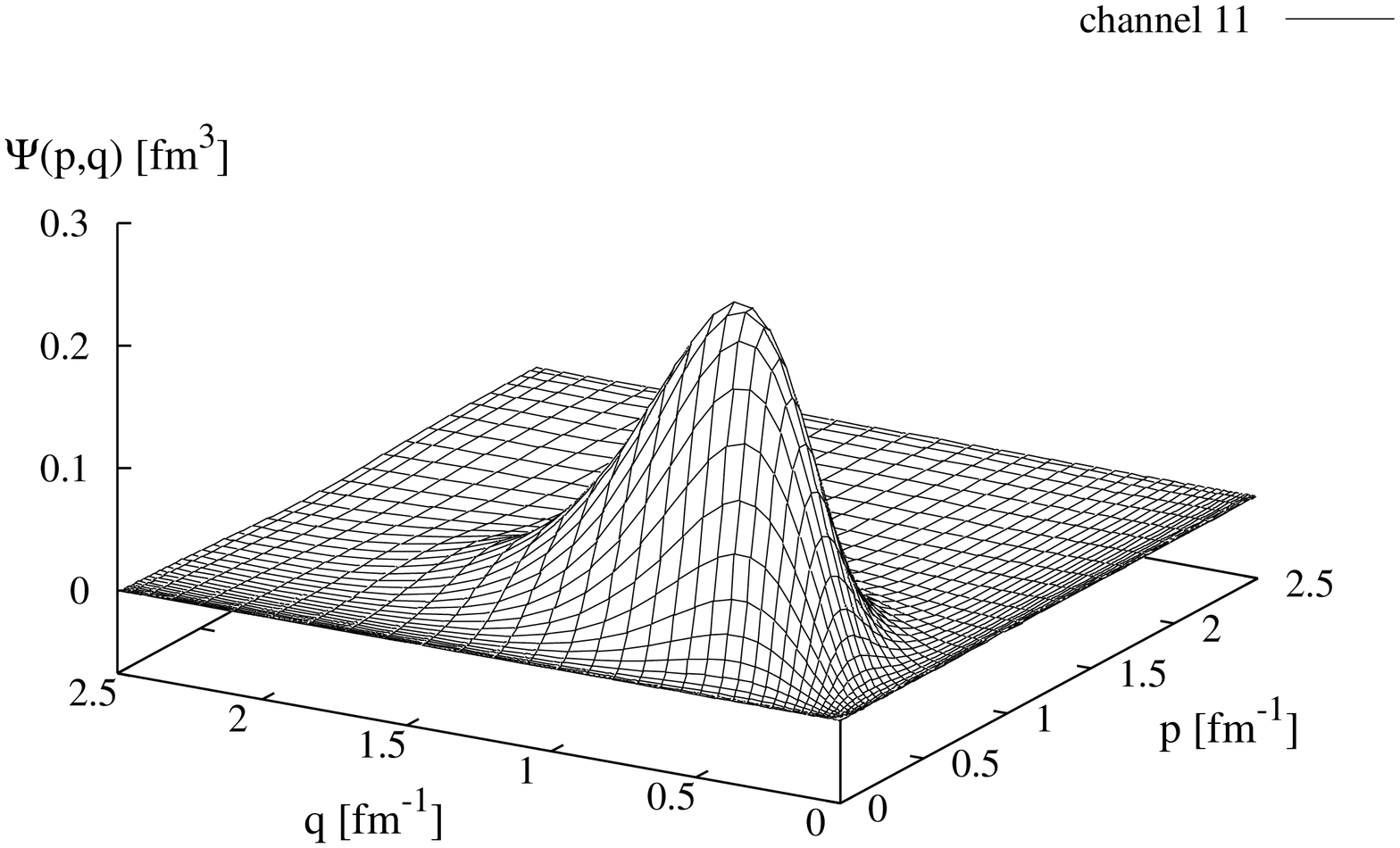,height=6.9cm,width=7.7cm,angle=-90}
\hspace{0.5cm}
\psfig{figure=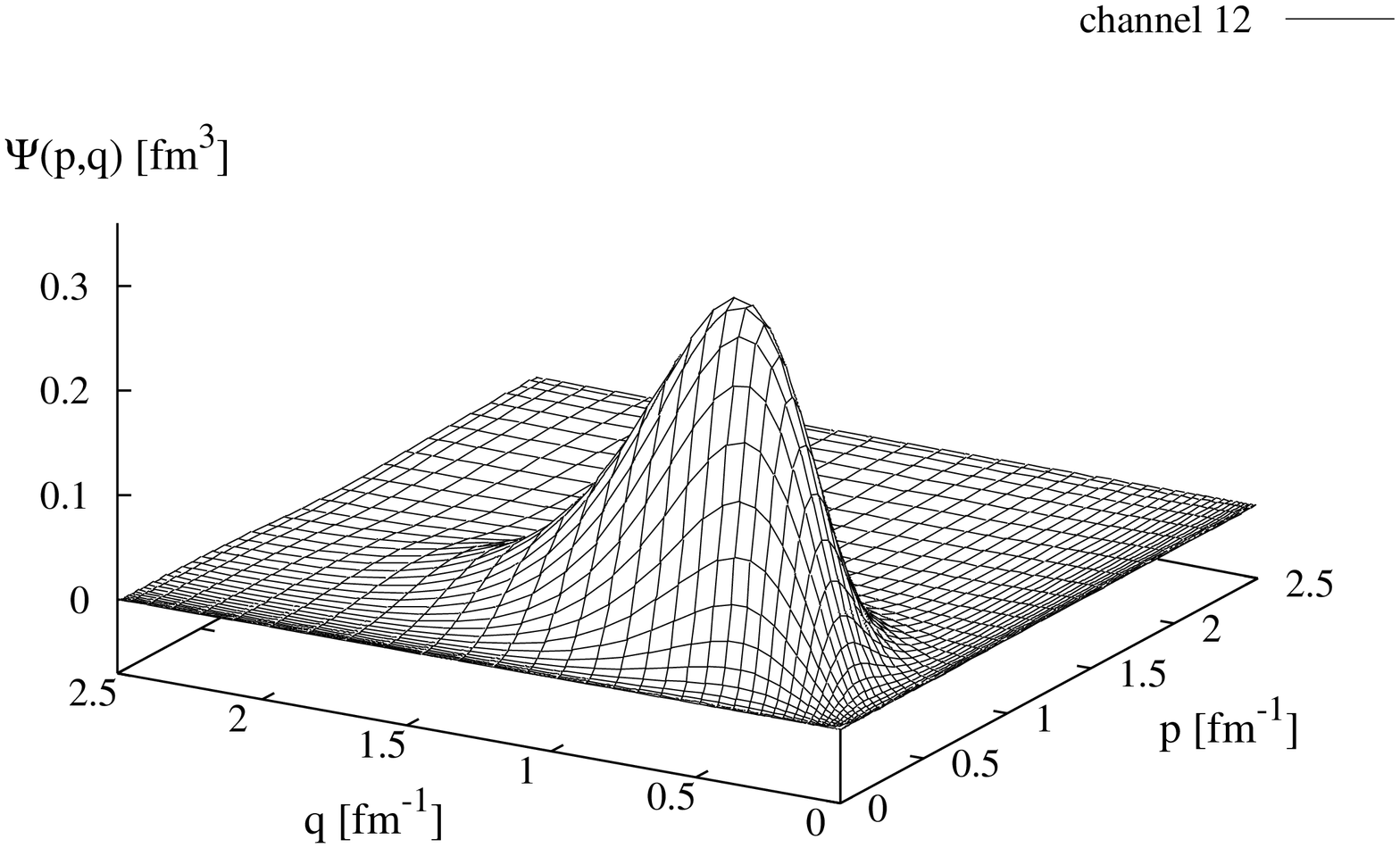,height=6.9cm,width=7.7cm,angle=-90}
}}
\caption{ -- continued.}
\end{figure}

\begin{figure}
\centerline{\hbox{
\psfig{figure=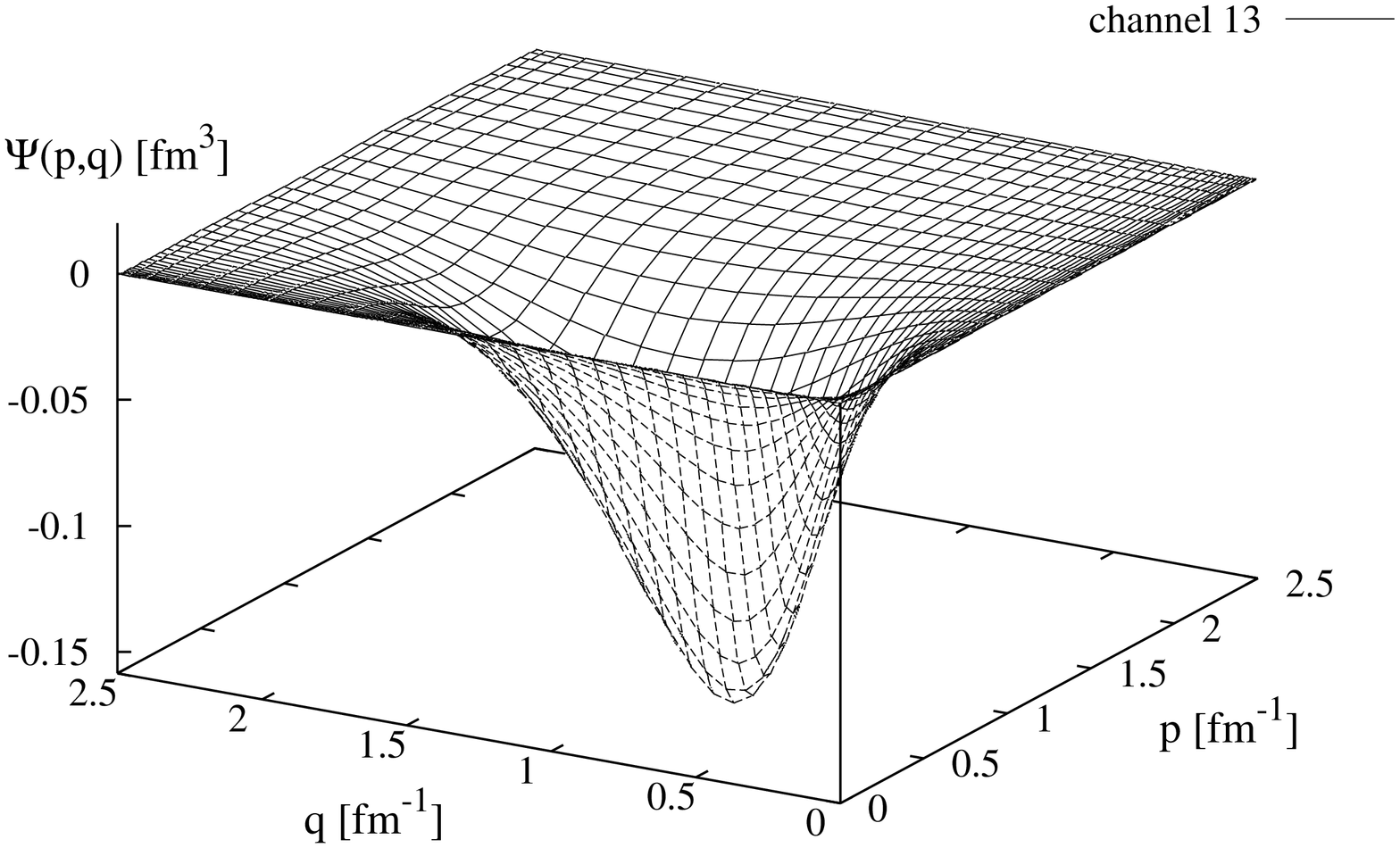,height=6.9cm,width=7.7cm,angle=-90}
\hspace{0.5cm}
\psfig{figure=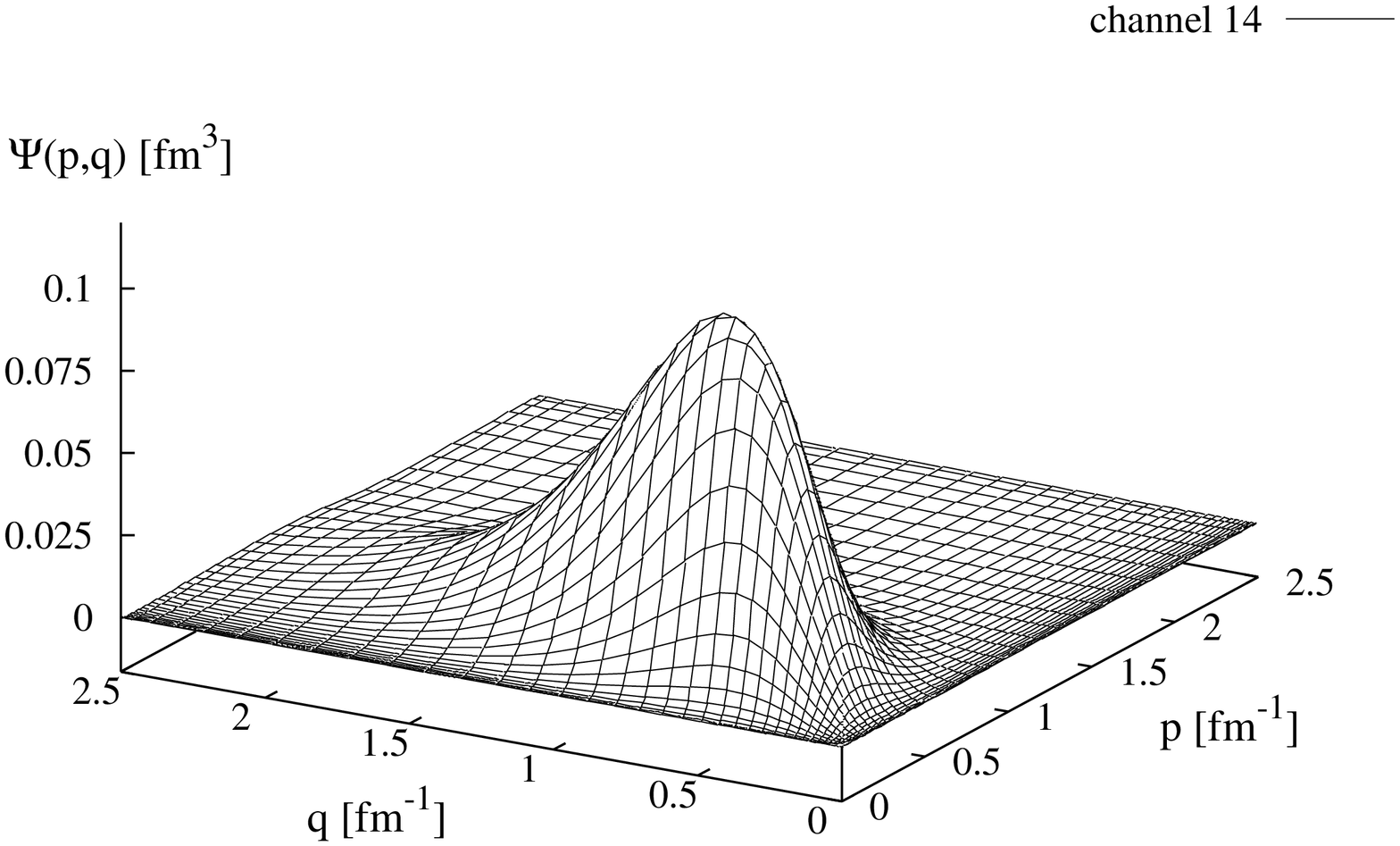,height=6.9cm,width=7.7cm,angle=-90}
}}
\centerline{\hbox{
\psfig{figure=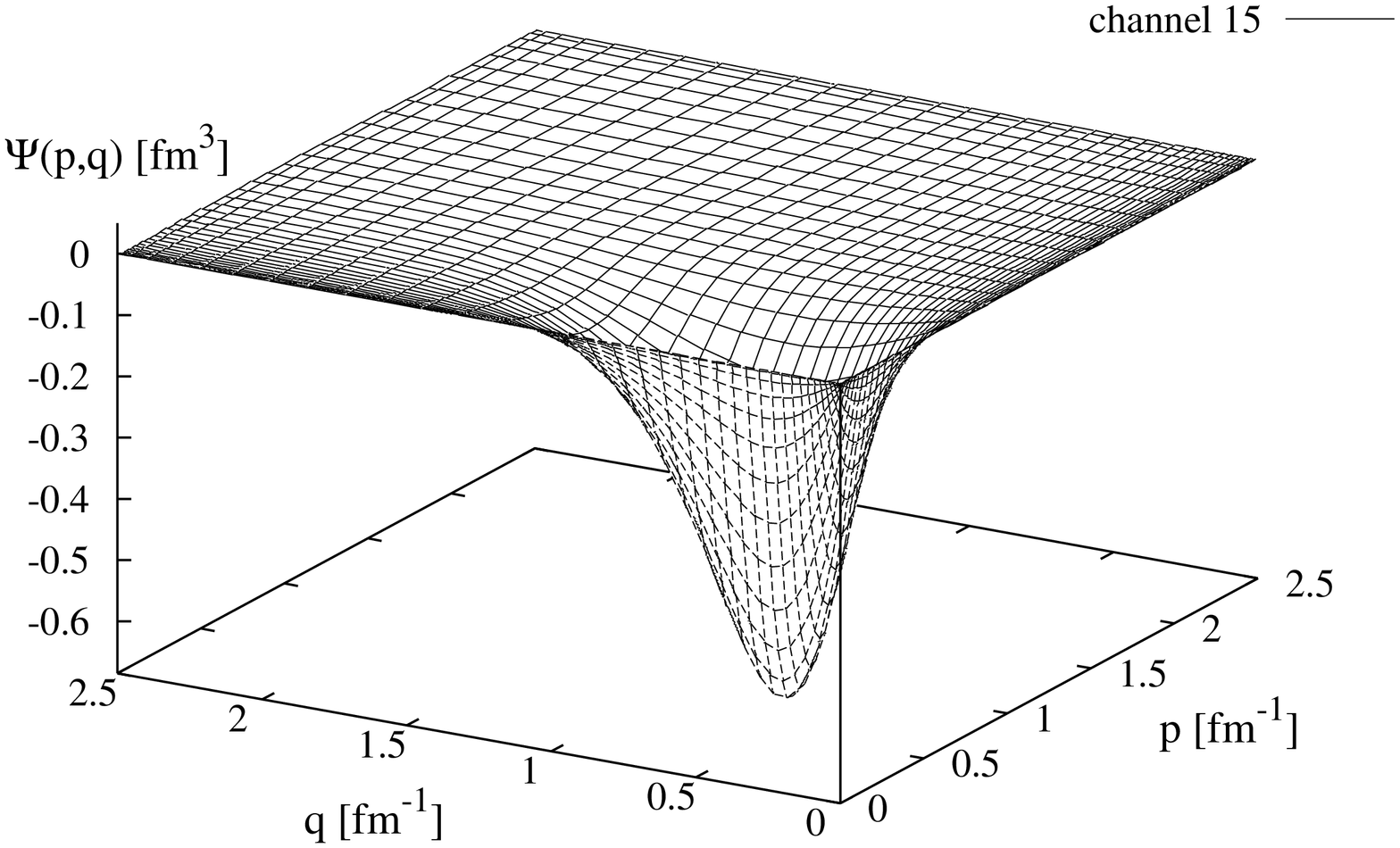,height=6.9cm,width=7.7cm,angle=-90}
\hspace{0.5cm}
\psfig{figure=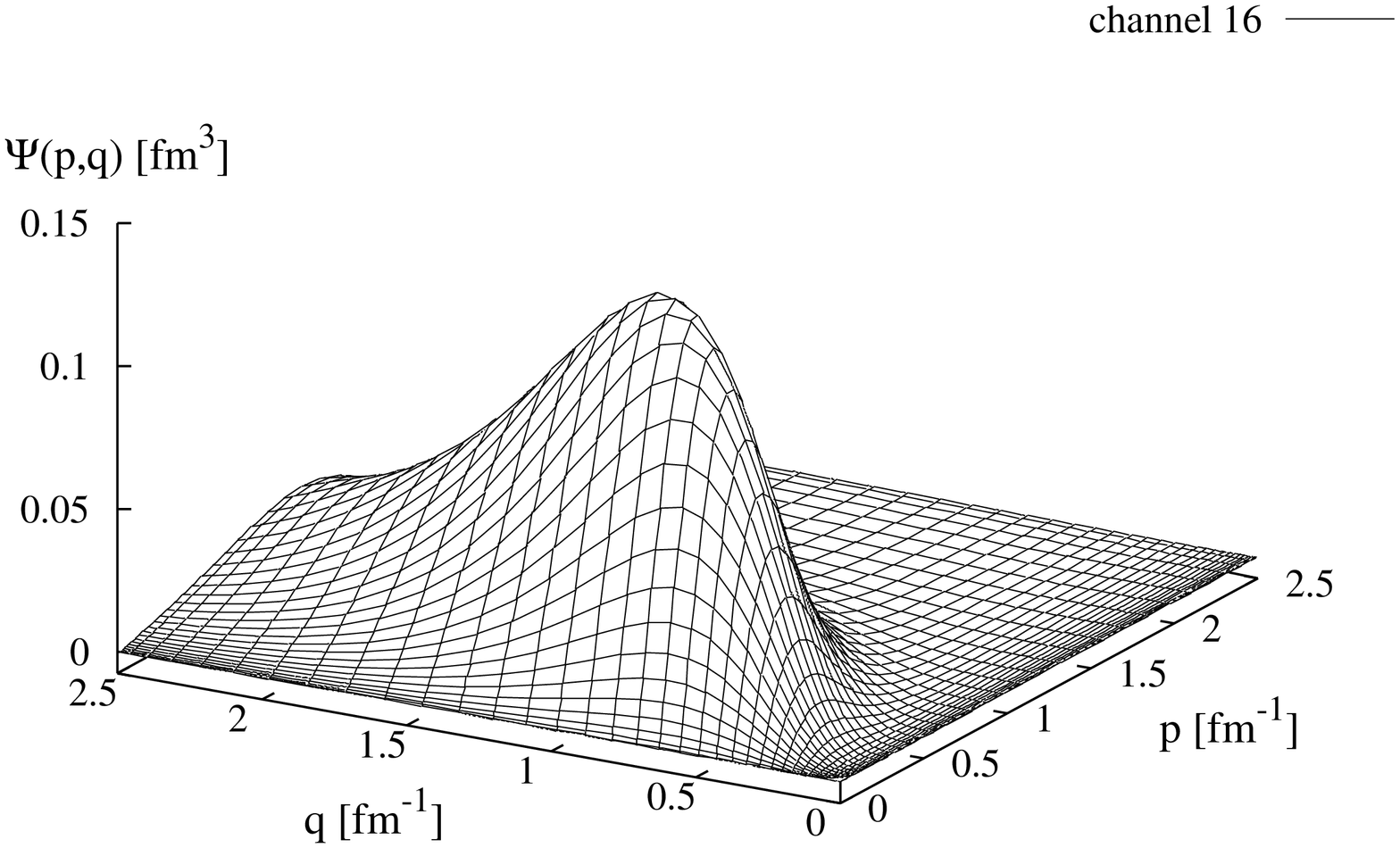,height=6.9cm,width=7.7cm,angle=-90}
}}
\centerline{\hbox{
\psfig{figure=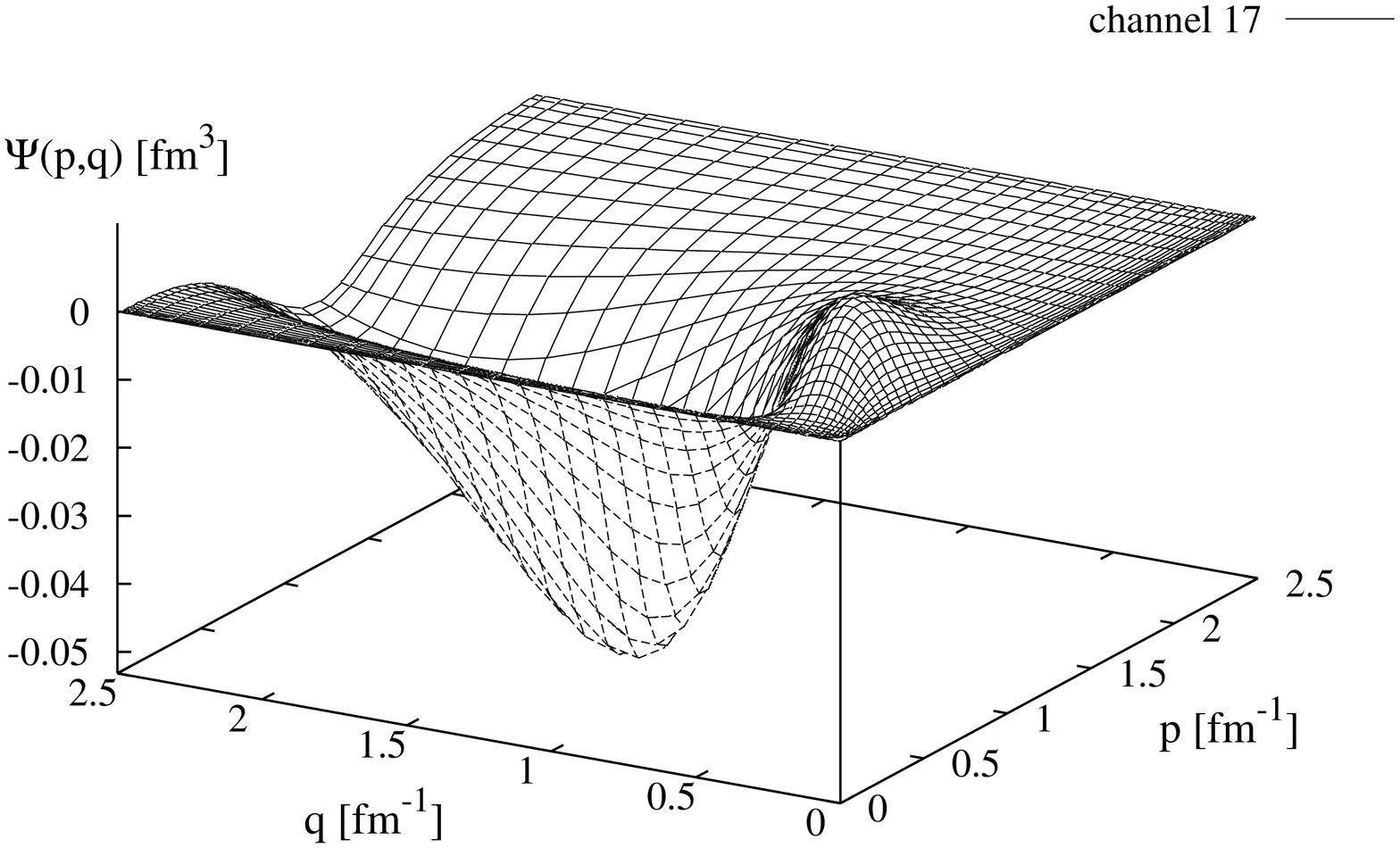,height=6.9cm,width=7.7cm,angle=-90}
\hspace{0.5cm}
\psfig{figure=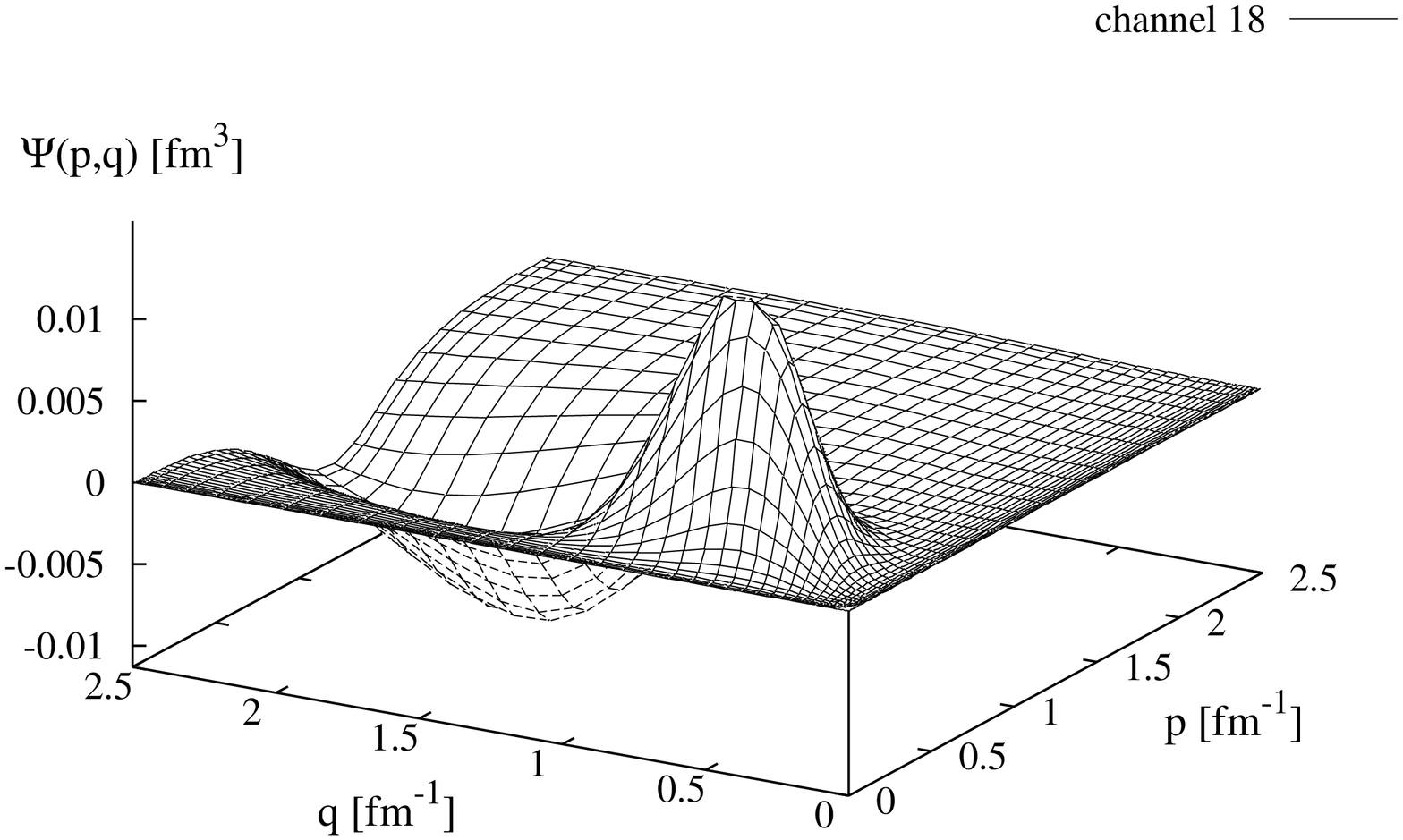,height=6.9cm,width=7.7cm,angle=-90}
}}
\caption{ -- continued.}
\end{figure}

\begin{figure}
\centerline{\hbox{
\psfig{figure=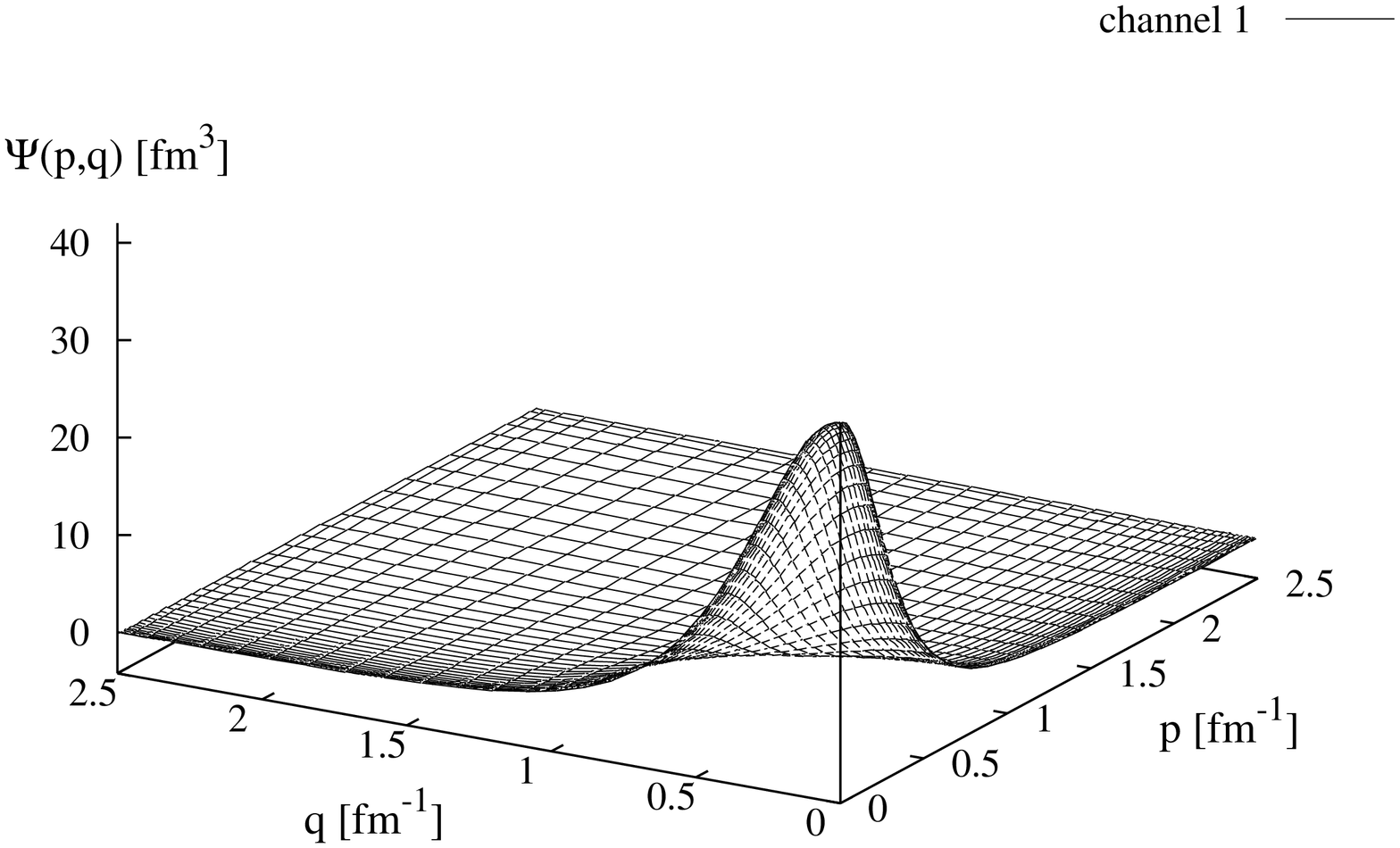,height=6.9cm,width=7.7cm,angle=-90}
\hspace{0.5cm}
\psfig{figure=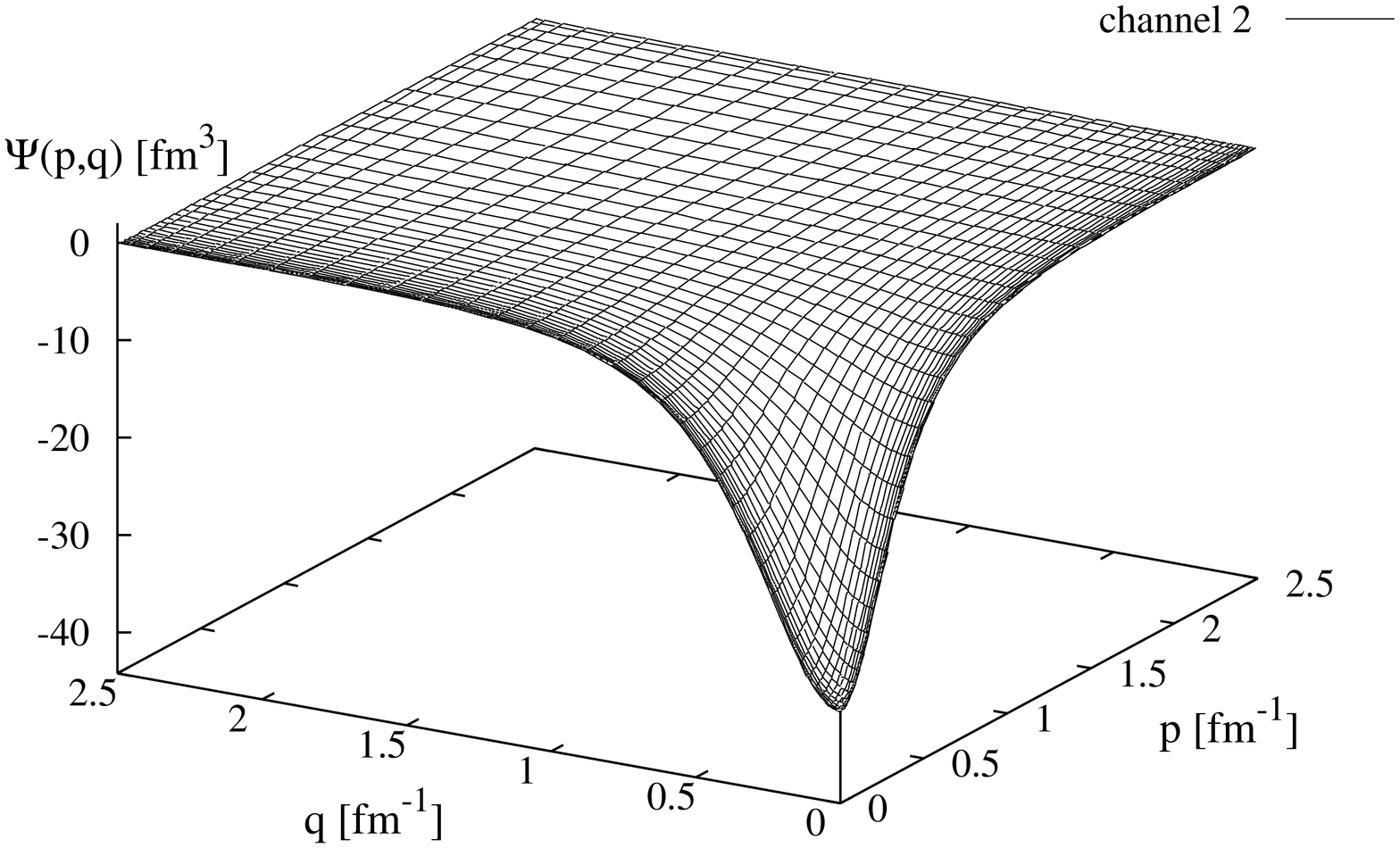,height=6.9cm,width=7.7cm,angle=-90}
}}
\centerline{\hbox{
\psfig{figure=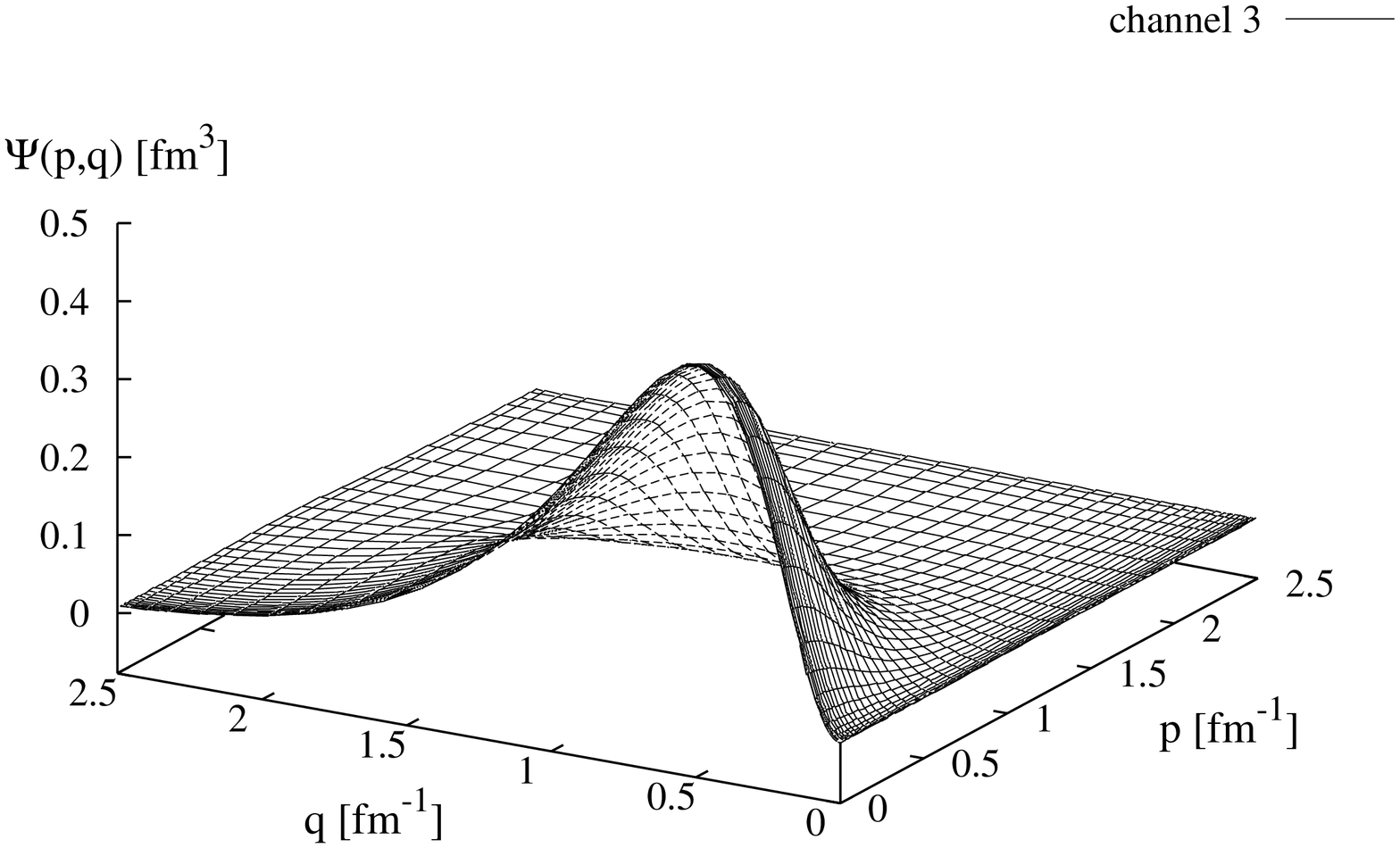,height=6.9cm,width=7.7cm,angle=-90}
\hspace{0.5cm}
\psfig{figure=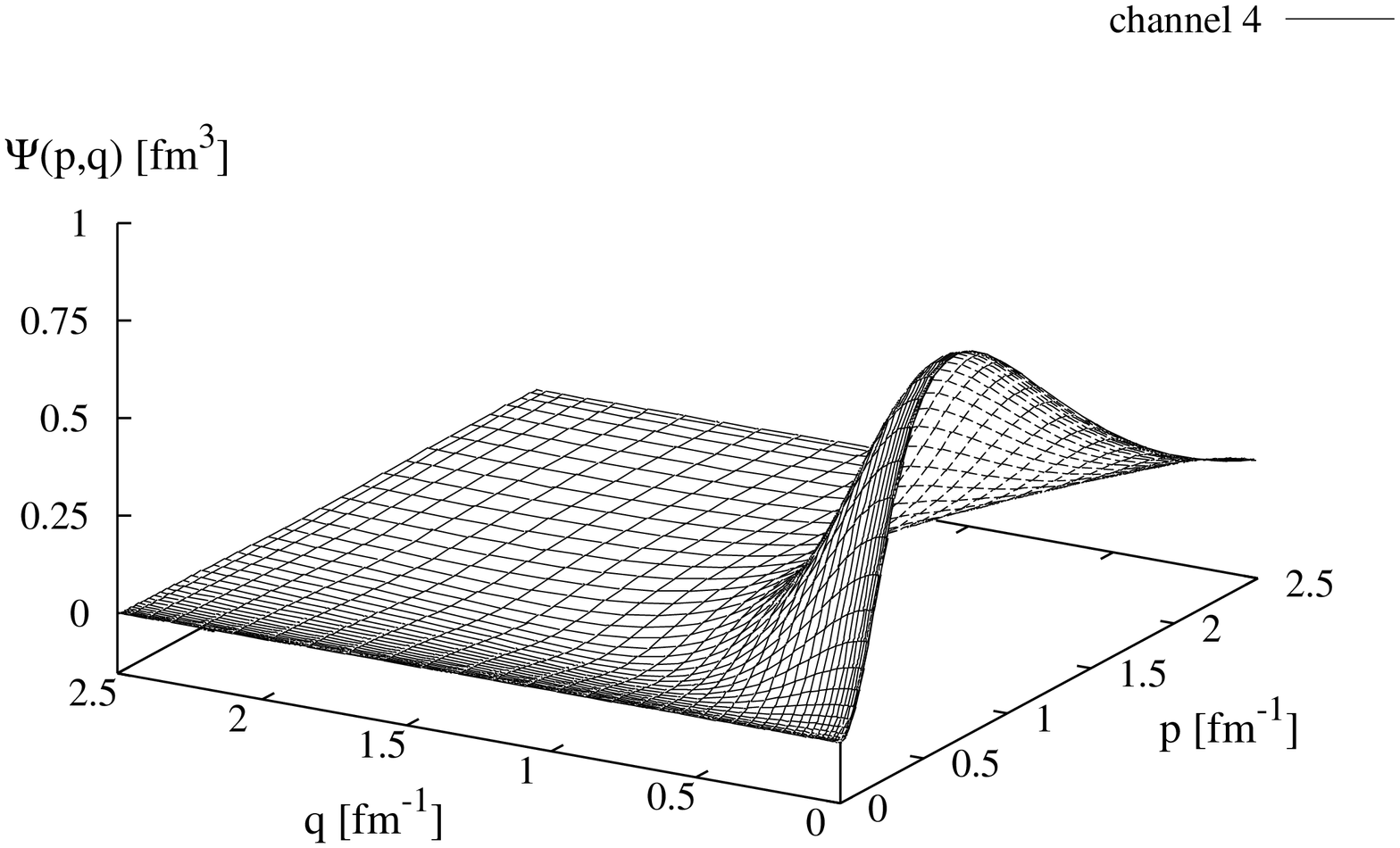,height=6.9cm,width=7.7cm,angle=-90}
}}
\centerline{\hbox{
\psfig{figure=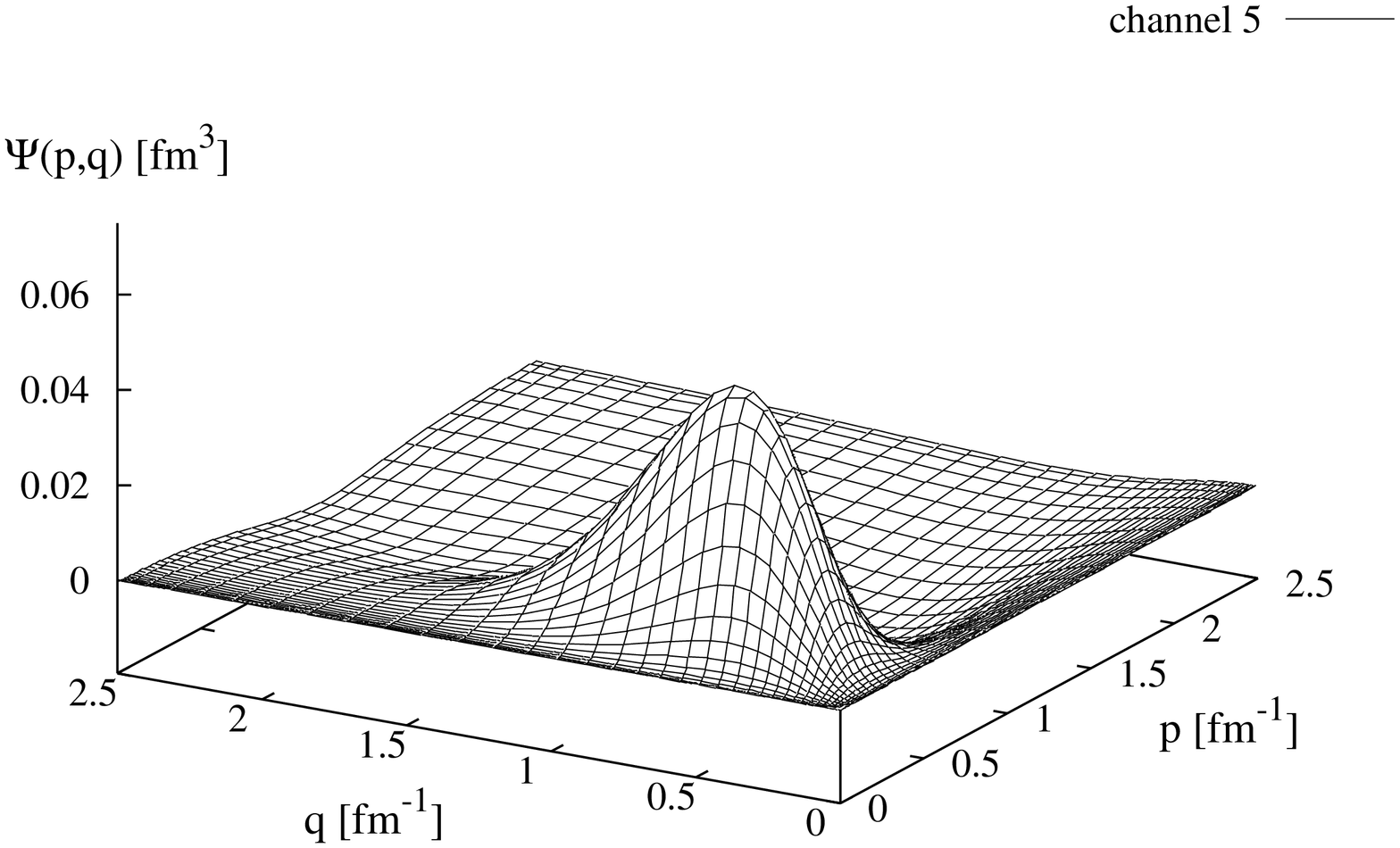,height=6.9cm,width=7.7cm,angle=-90}
\hspace{0.5cm}
\psfig{figure=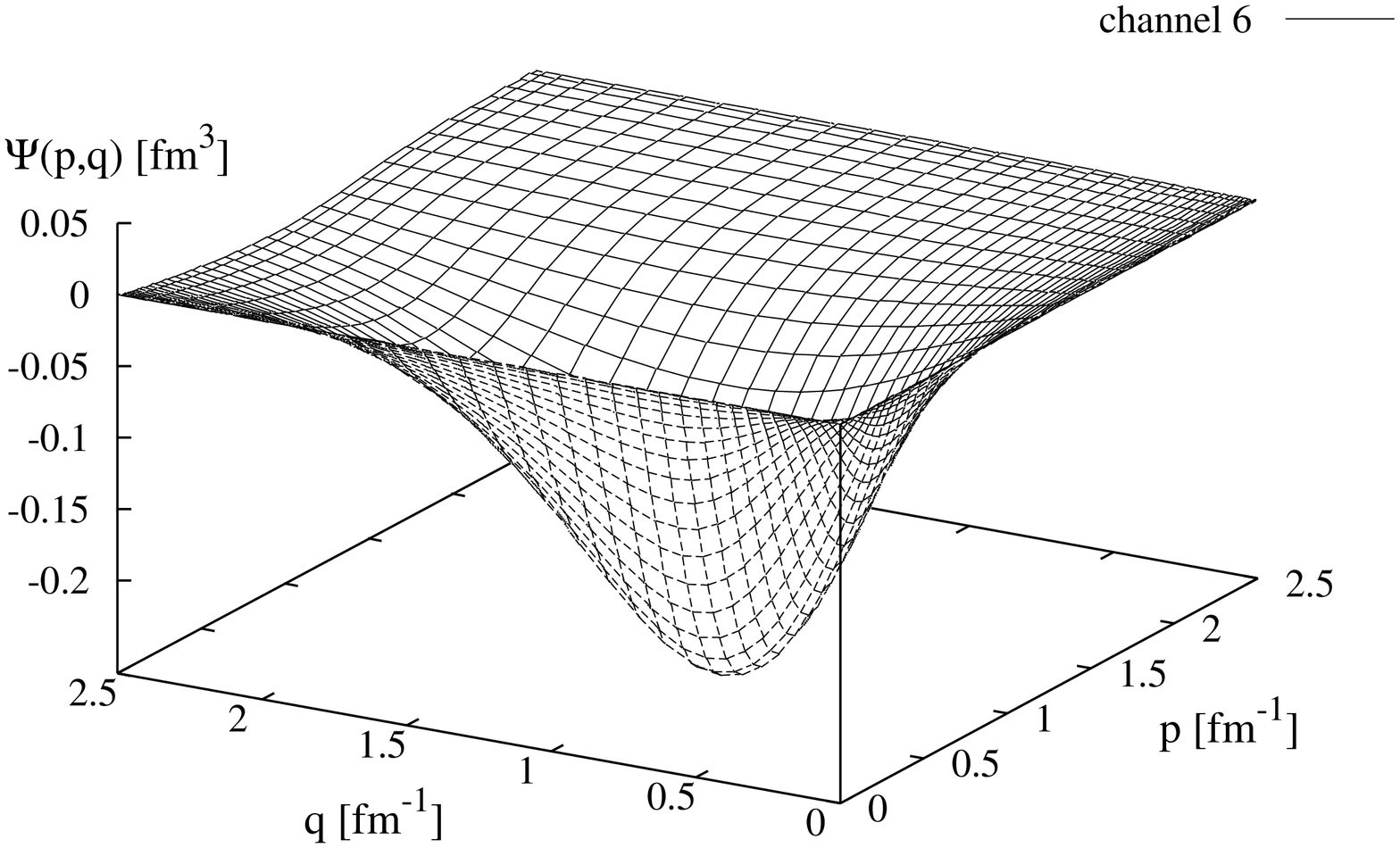,height=6.9cm,width=7.7cm,angle=-90}
}}
\caption{Same as Fig. 1 but for the Bonn {\sl A} (EST) potential.}
\end{figure}

\begin{figure}
\centerline{\hbox{
\psfig{figure=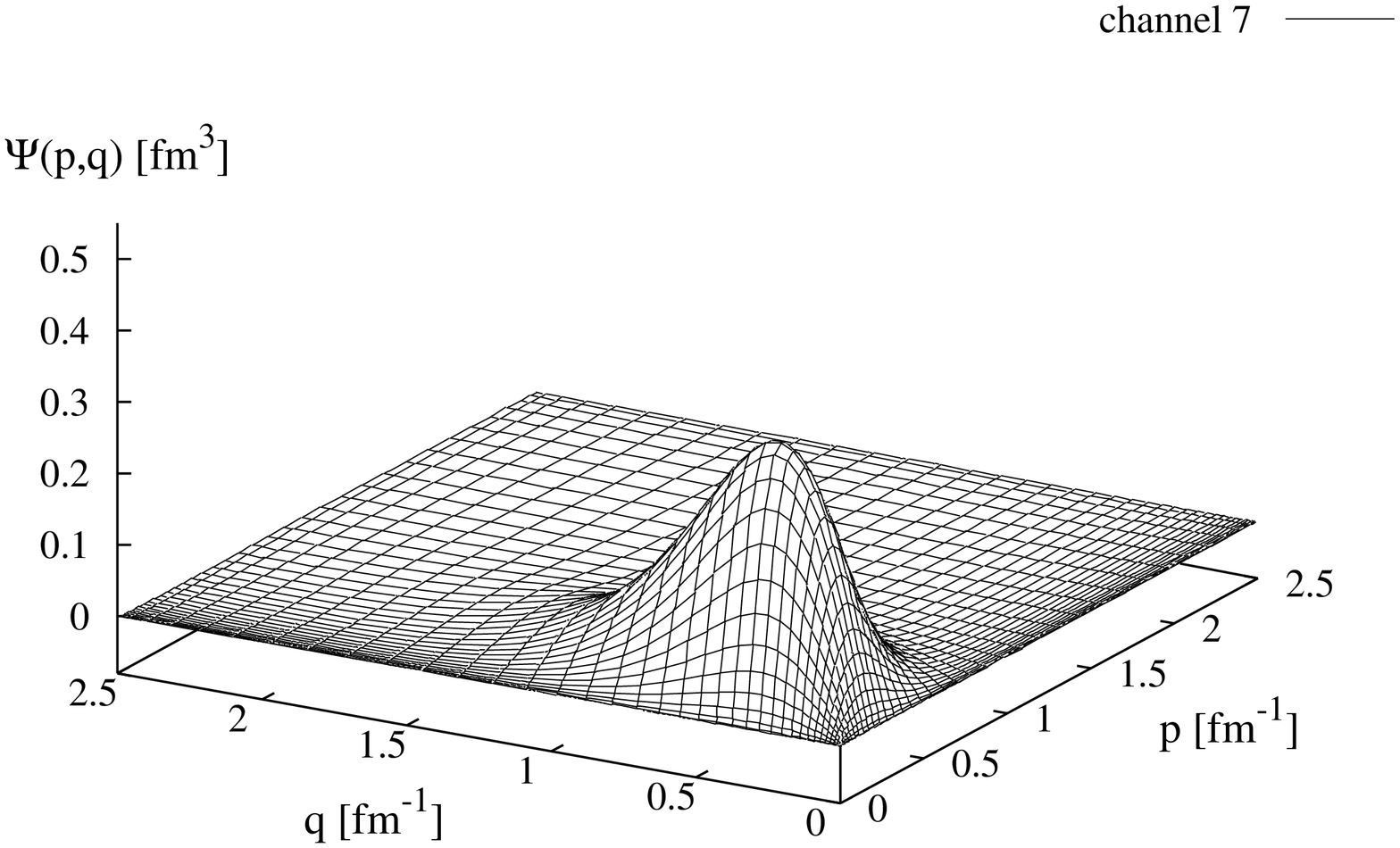,height=6.9cm,width=7.7cm,angle=-90}
\hspace{0.5cm}
\psfig{figure=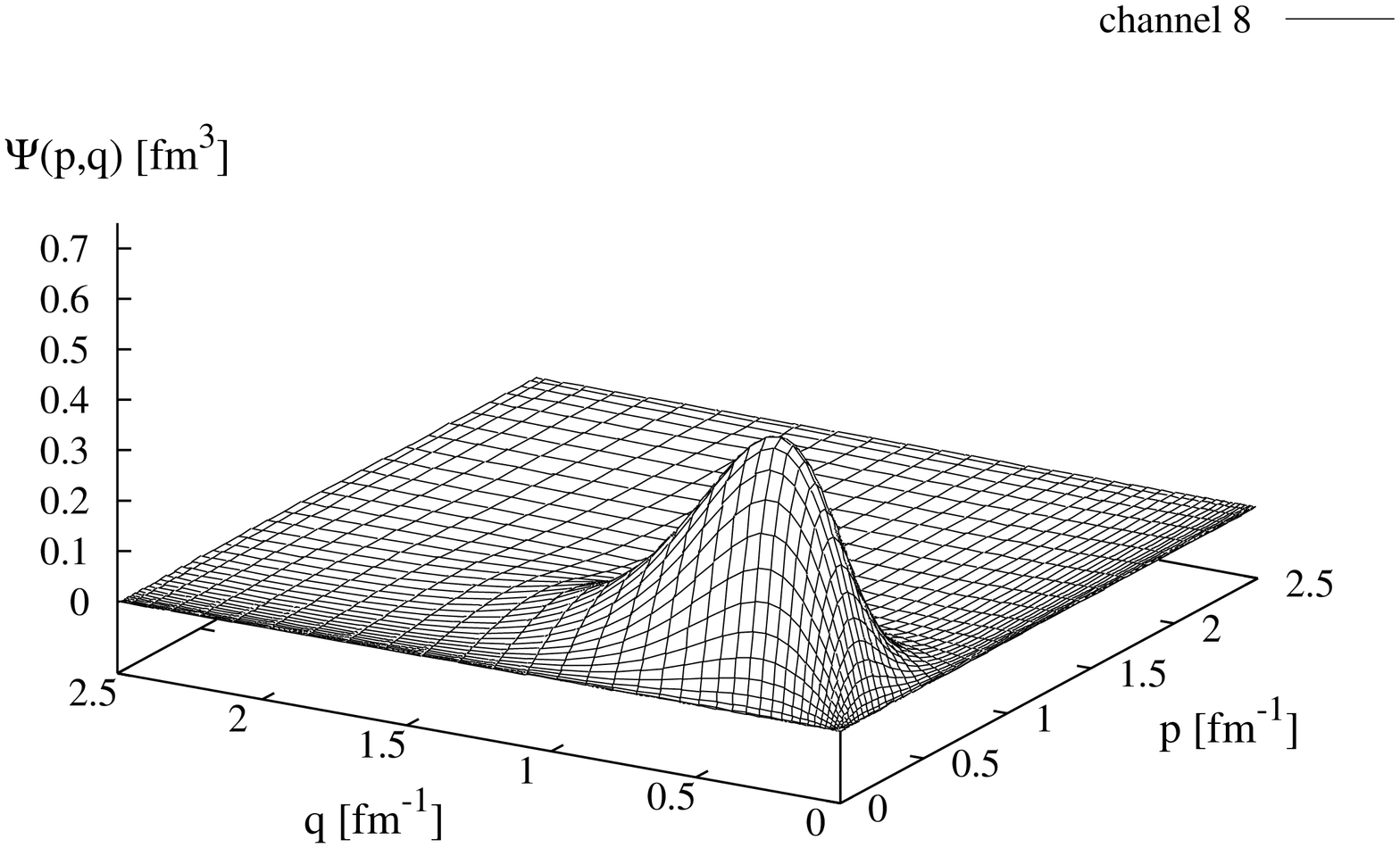,height=6.9cm,width=7.7cm,angle=-90}
}}
\centerline{\hbox{
\psfig{figure=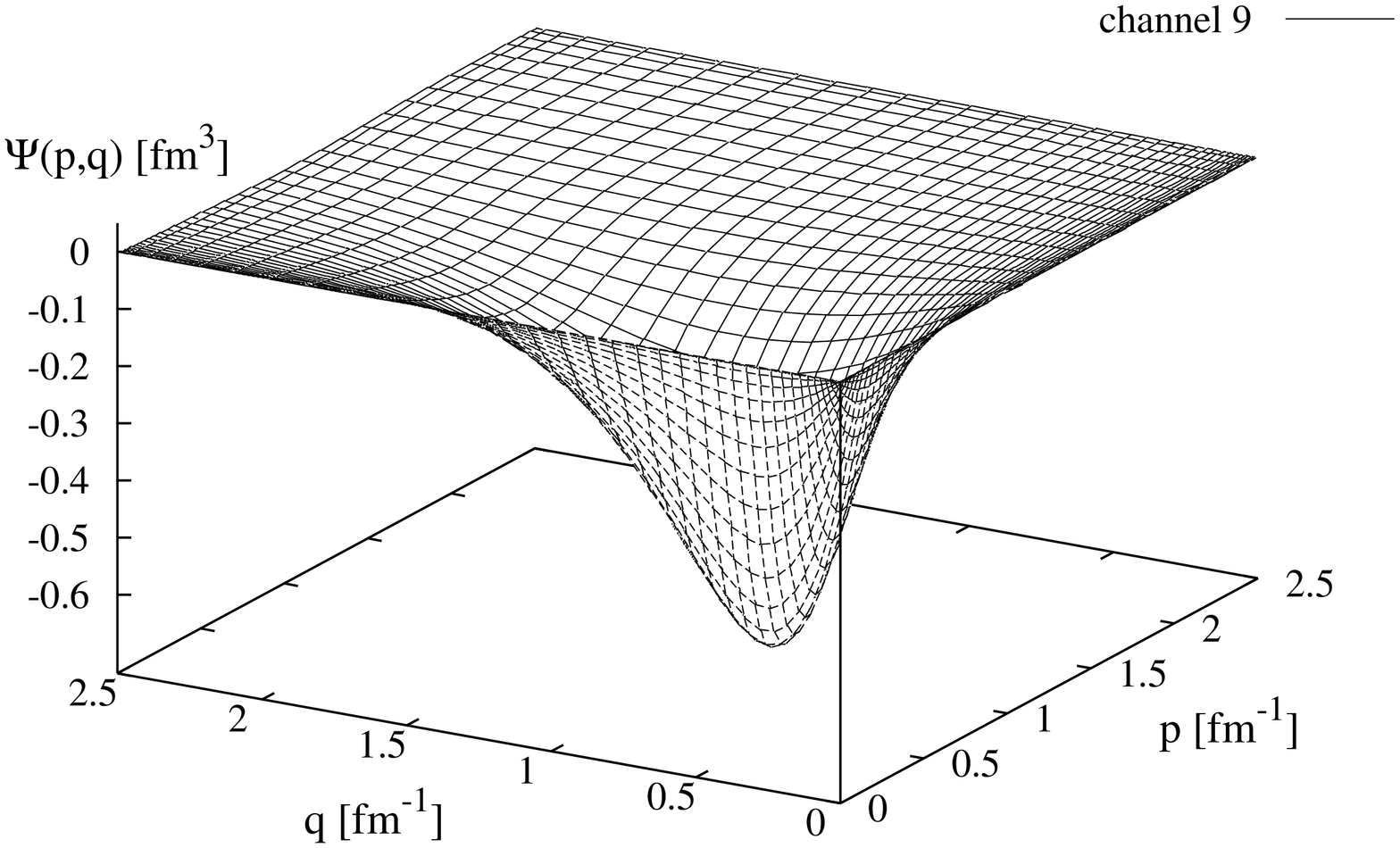,height=6.9cm,width=7.7cm,angle=-90}
\hspace{0.5cm}
\psfig{figure=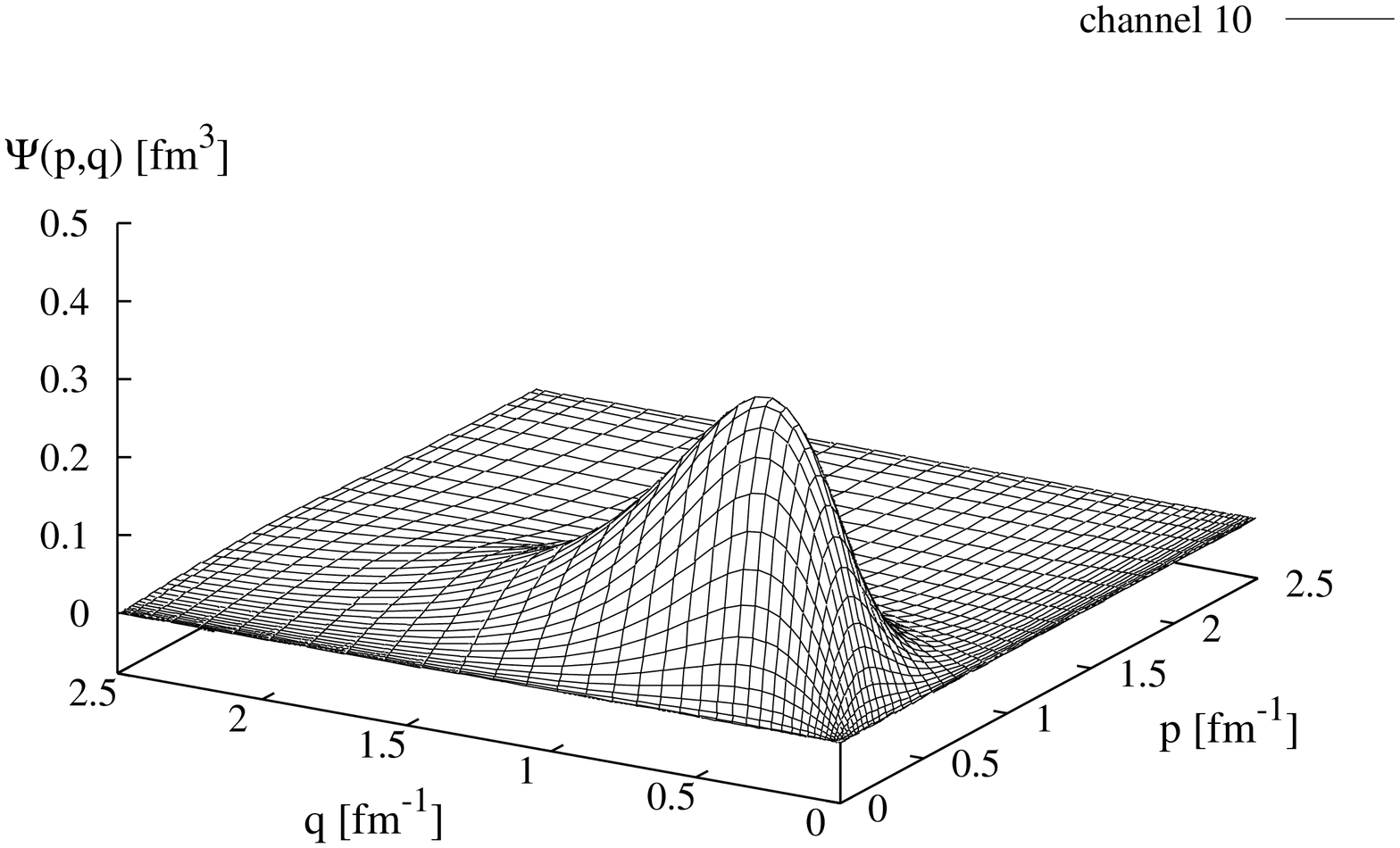,height=6.9cm,width=7.7cm,angle=-90}
}}
\centerline{\hbox{
\psfig{figure=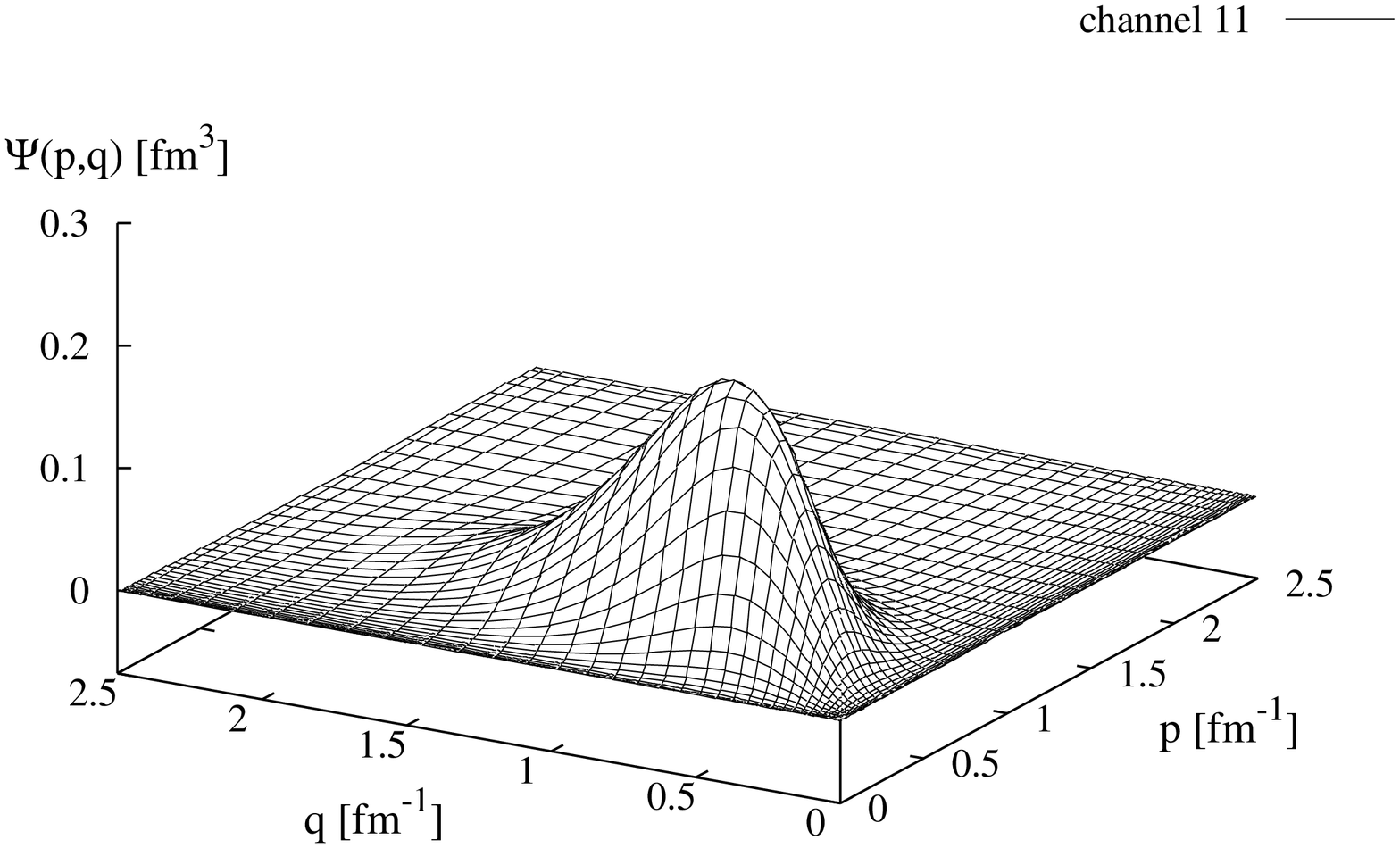,height=6.9cm,width=7.7cm,angle=-90}
\hspace{0.5cm}
\psfig{figure=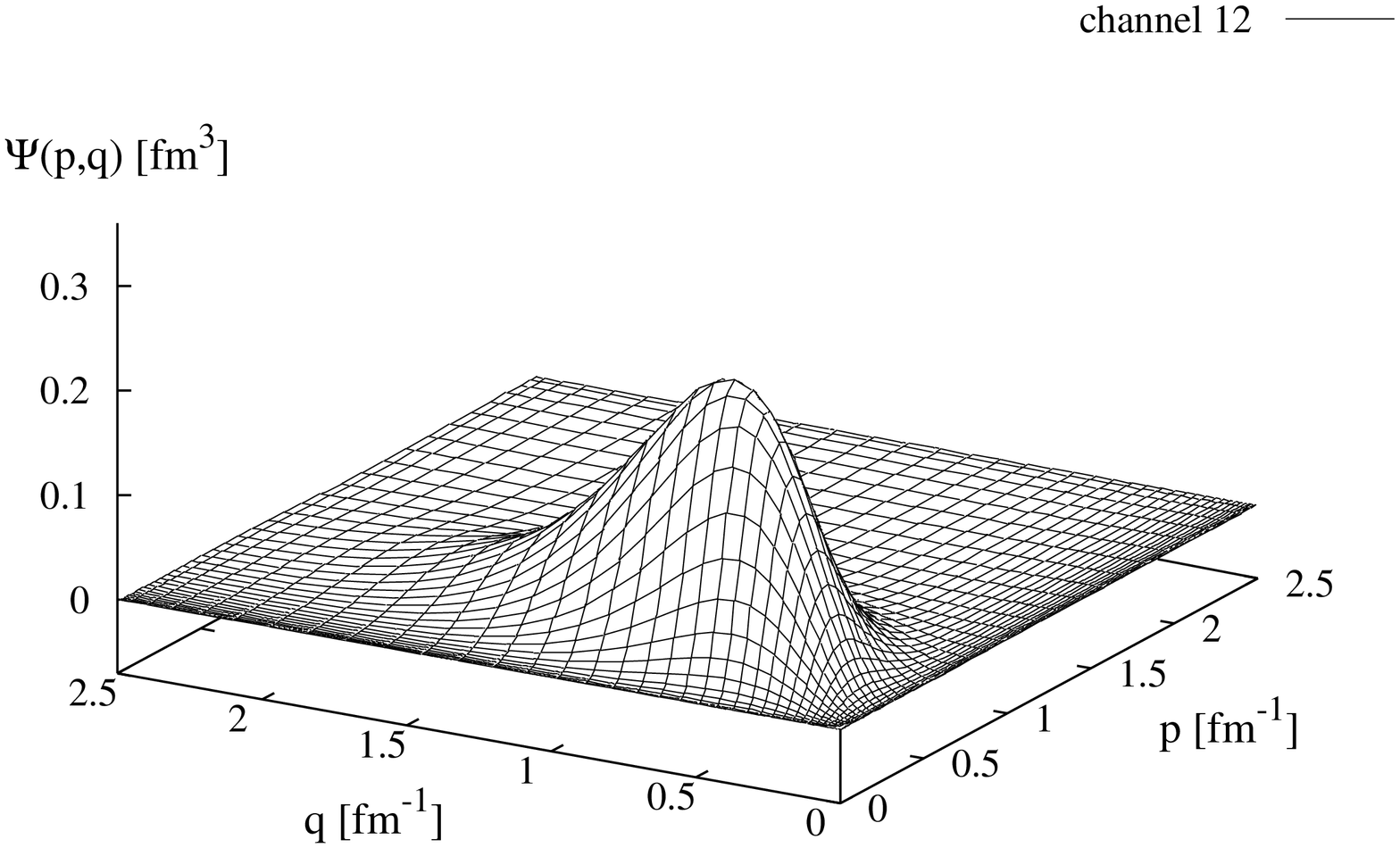,height=6.9cm,width=7.7cm,angle=-90}
}}
\caption{ -- continued.}
\end{figure}

\begin{figure}
\centerline{\hbox{
\psfig{figure=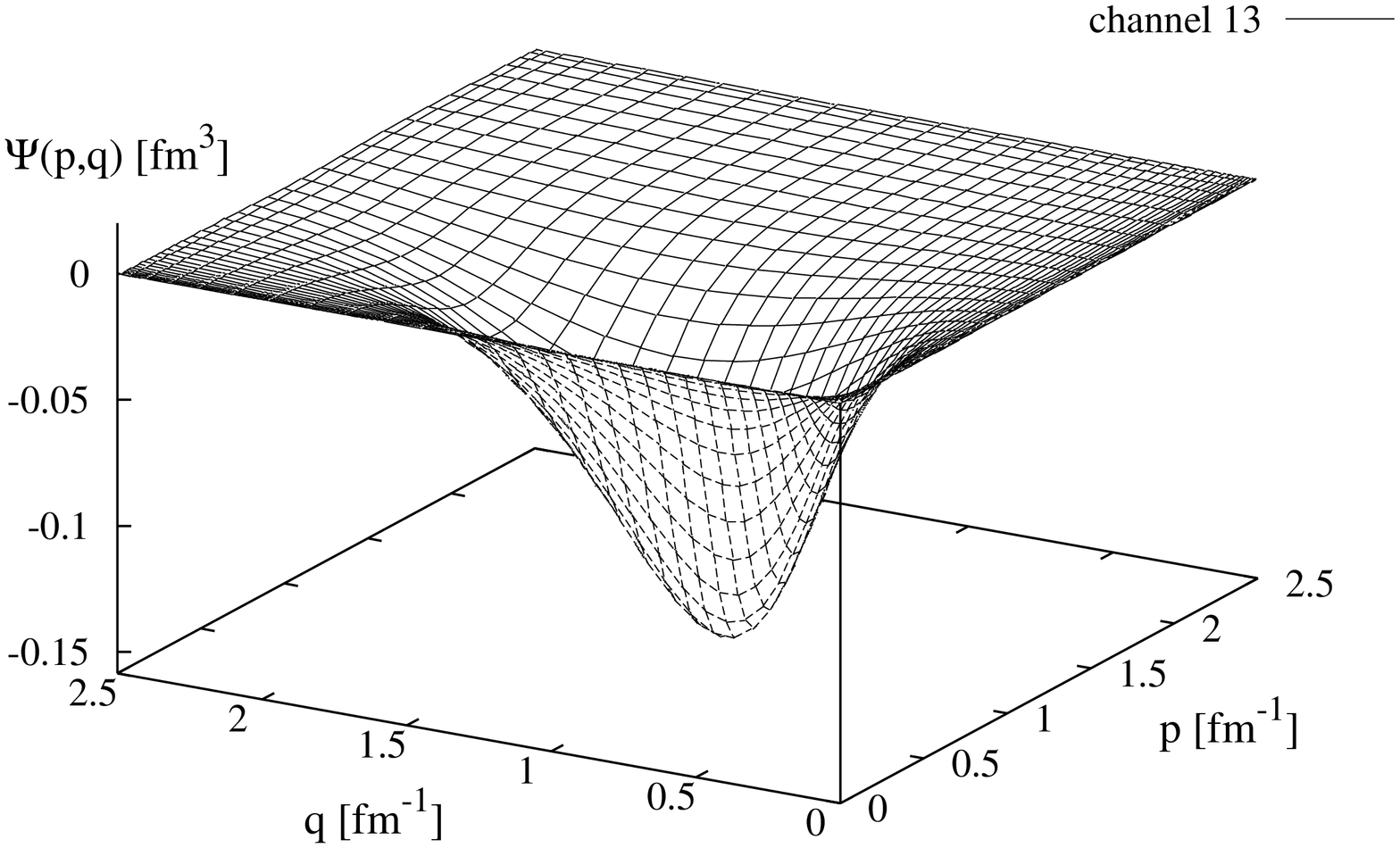,height=6.9cm,width=7.7cm,angle=-90}
\hspace{0.5cm}
\psfig{figure=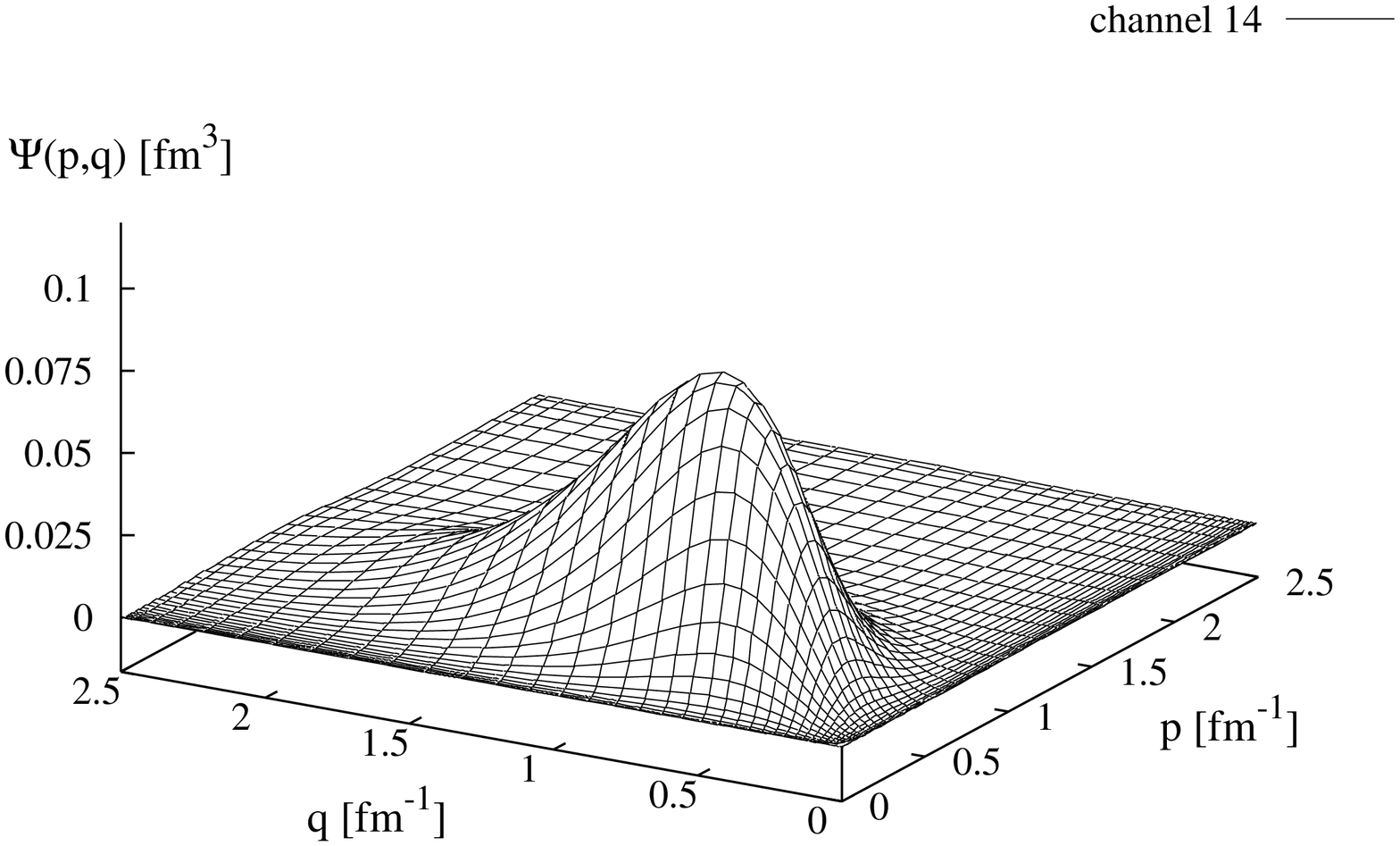,height=6.9cm,width=7.7cm,angle=-90}
}}
\centerline{\hbox{
\psfig{figure=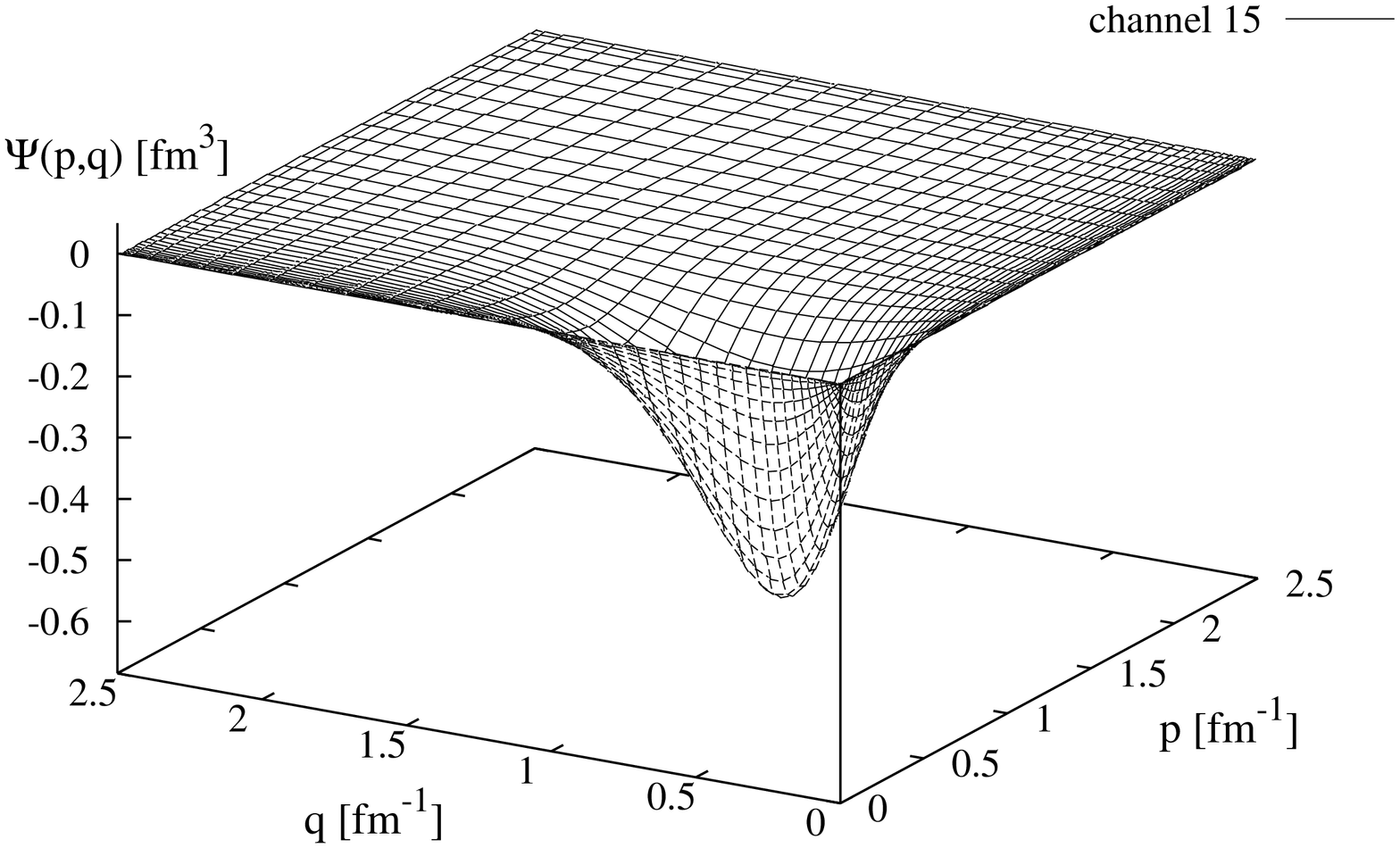,height=6.9cm,width=7.7cm,angle=-90}
\hspace{0.5cm}
\psfig{figure=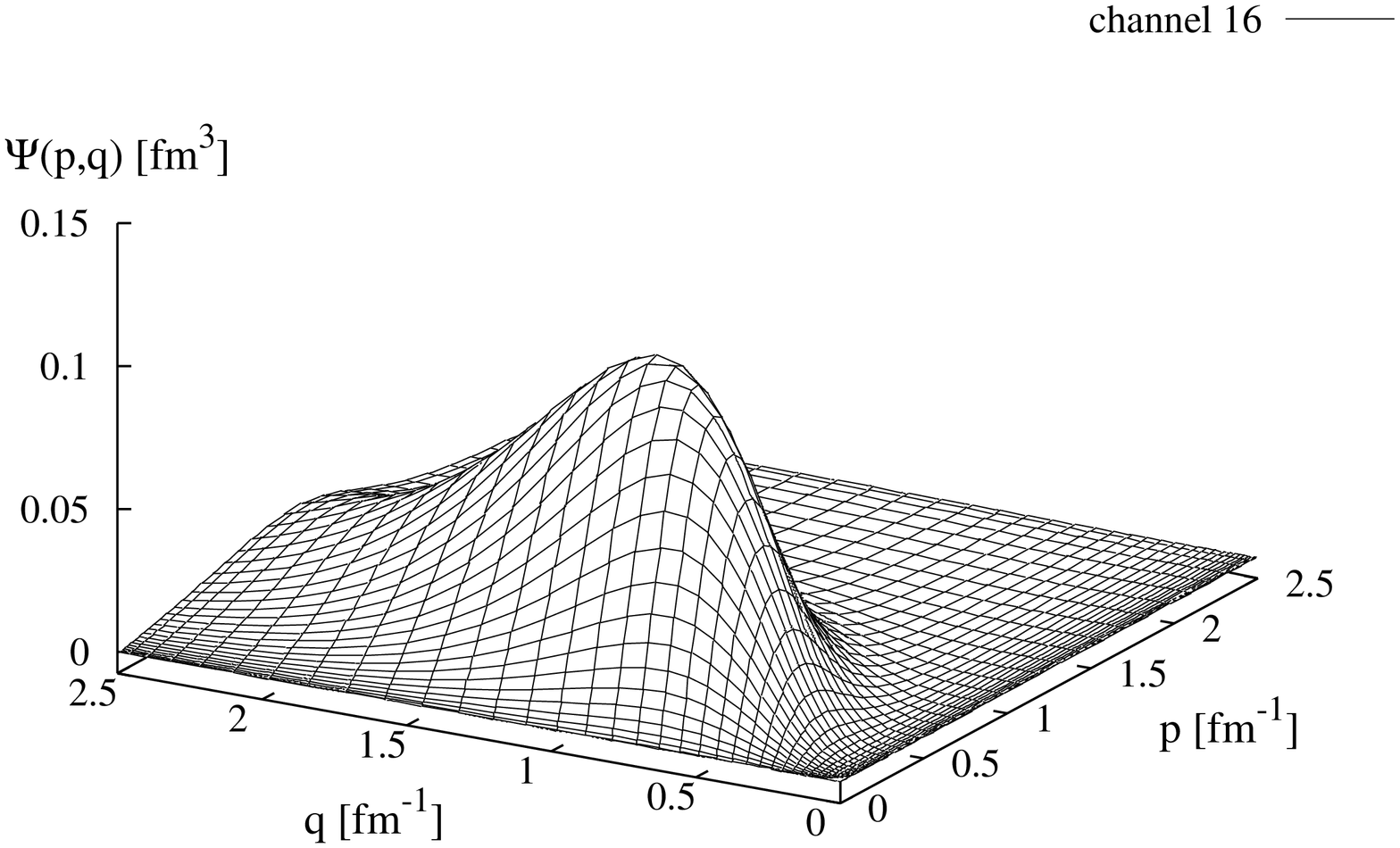,height=6.9cm,width=7.7cm,angle=-90}
}}
\centerline{\hbox{
\psfig{figure=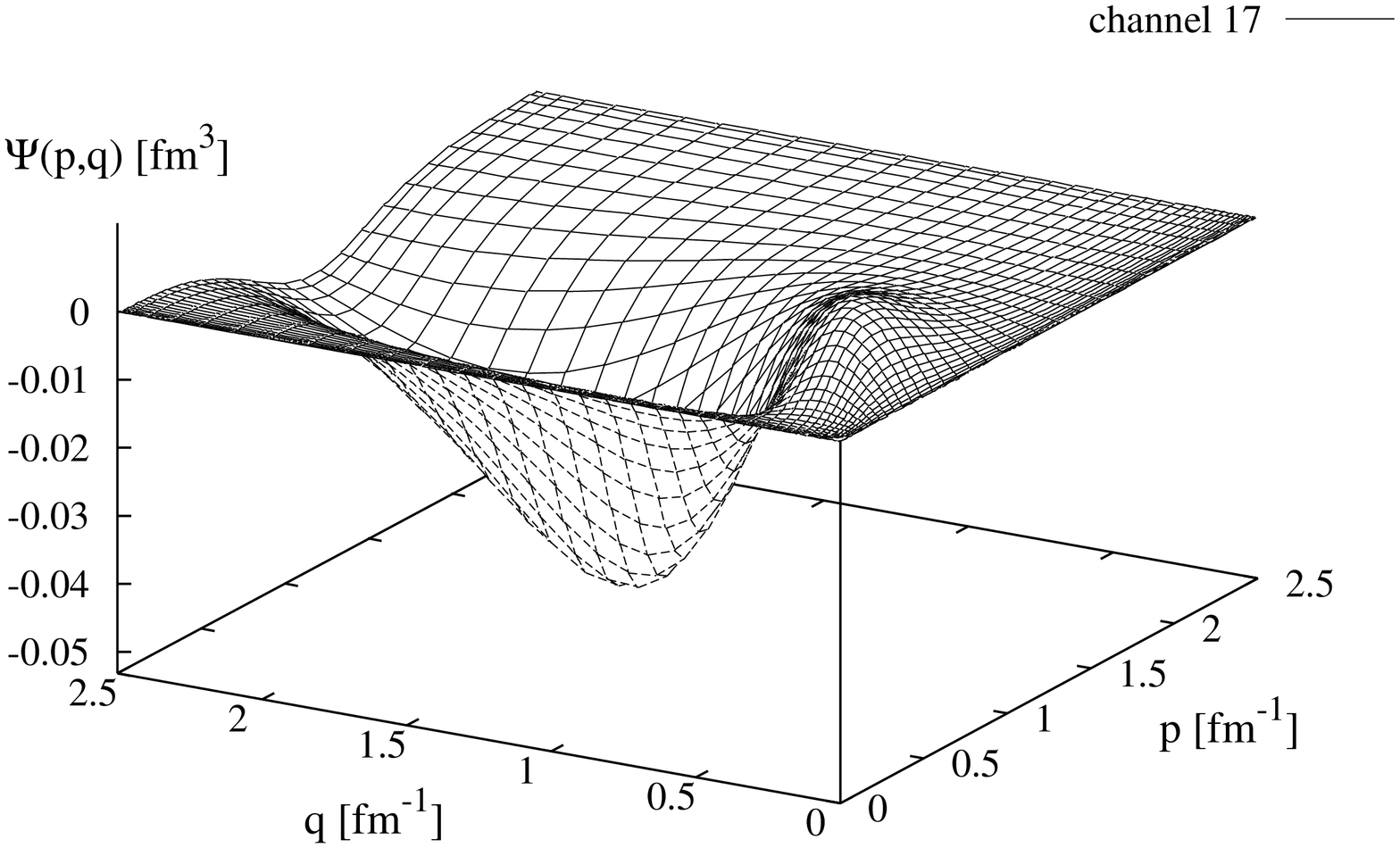,height=6.9cm,width=7.7cm,angle=-90}
\hspace{0.5cm}
\psfig{figure=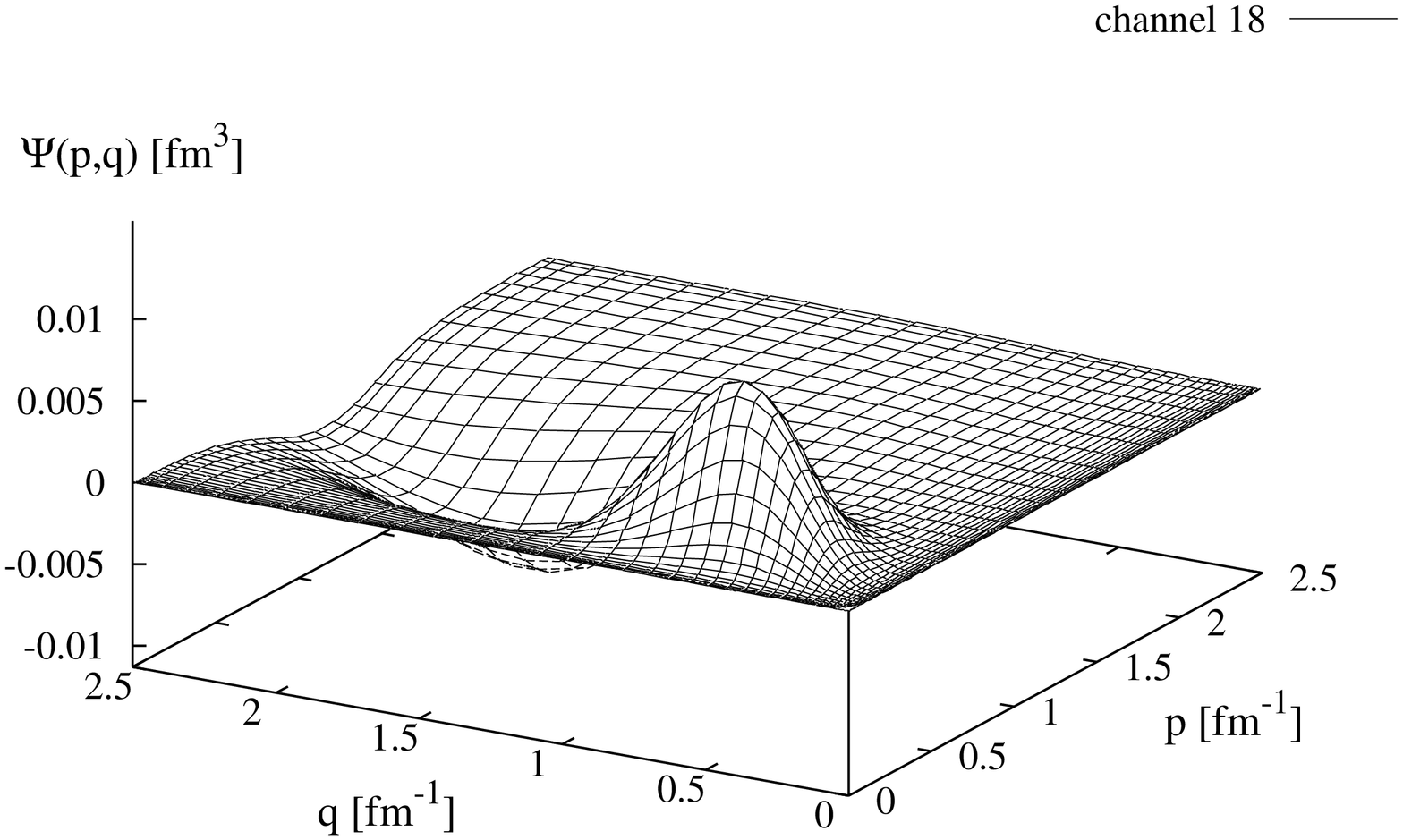,height=6.9cm,width=7.7cm,angle=-90}
}}
\caption{ -- continued.}
\end{figure}

\end{document}